\documentclass[twocolumn,showpacs,preprintnumbers,amsmath,amssymb]{revtex4}

\usepackage[unicode]{hyperref}
\usepackage[caption=false]{subfig}
\usepackage{graphics}
\usepackage{epstopdf}
\usepackage{epsfig}
\usepackage{graphicx}
\graphicspath{ {./images/} }
\usepackage{dcolumn}
\usepackage{bm}
\usepackage[caption=false]{subfig}

\begin{document}

\title{Universality class of site and bond percolation on multi-multifractal scale-free planar stochastic lattice
}%

\author{M. K. Hassan and M. M. Rahman}%
\date{\today}%

\affiliation{
University of Dhaka, Department of Physics, Dhaka 1000, Bangladesh \\
}

\begin{abstract}

In this article, we investigate both site and bond percolation on a weighted planar stochastic lattice (WPSL) which is a multi-multifractal
and whose dual is a scale-free network. The characteristic properties of percolation is that it exhibits threshold phenomena as we find sudden
or abrupt jump in spanning probability across $p_c$ accompanied by the divergence of some other observable quantities which is reminiscent of continuous phase transition.
Indeed, percolation is characterized by the critical behavior of percolation strength $P(p)\sim (p_c-p)^\beta$, mean cluster size $S\sim (p_c-p)^{-\gamma}$ and the 
system size $L\sim (p_c-p)^{-\nu}$ which are known as the equivalent counterpart of the order parameter, susceptibility and correlation length respectively. Moreover, 
the cluster size distribution function $n_s(p_c)\sim s^{-\tau}$ and the mass-length relation $M\sim L^{d_f}$ of the spanning cluster also 
provide useful characterization of the percolation process. We obtain an exact value for $p_c$ and for all the exponents such as $\beta, \nu, \gamma, \tau$ and $d_f$. We find
that, except $p_c$, all the exponents are exactly the same in both bond and site percolation despite the significant difference in the definition of cluster
and other quantities. Our results suggest that the percolation on WPSL belongs to a new universality class as its exponents do not share the same value as for 
all the existing planar lattices and like other cases its site and bond belong to the same universality class.

\end{abstract}


\maketitle

\section{Introduction}

Percolation is perhaps one of the most studied problems in statistical physics. This is not only 
because of the simplicity of its definition but also because of the versatility of its applications.
To study percolation one needs to choose a skeleton first. It can be a lattice or a graph
that has two entities namely sites (nodes) and bonds (edges). We then occupy each site or bond,
depending on whether we want to study site or bond percolation, 
with probability $p$ independent of the state of its neighbors \citep{ref.Stauffer,ref.bunde}. 
Broadbent and Hammersley in 1957 first presented the percolation
model to understand the motion of gas molecules through the maze of pores in carbon granules 
filling a gas mask \cite{ref.broadbent}. Since then the intuitive idea of percolation 
has been found relevant to so many seemingly disperate systems that its concept has literally percolated 
across a vast area of science and social science. Examples include flow of fluid in porous media,
infiltration in composite materials processing, spread of fluids, rumours, opinion, biological and 
computer viruses are just a few to mention  \citep {ref.Dashtian, ref.Gennes, ref.boccaletti_opinion, ref.mendes_rumor,ref.pastor_rumor,
ref.Newman_virus,ref.Moore_virus,ref.Cohen_virus}. 

Besides the simplicity of its definition and the versatility of its application there exists
 yet another reason why percolation model is so popular. In percolation we 
primarily observe how clusters, set of contiguous occupied sites, are formed and grown
as a function of $p$ which is the only control parameter.
As $p$ value increases from negligibly small,  there appears for the first time a cluster 
that spans across the entire system. In the case of infinite system size,
we find a unique threshold value $p_c$ such that there is the probability that the spanning cluster $W(p)=0$
for $p\leq p_c$ and $W(p)=1$ for $p>p_c$.
Interestingly, such transition, despite being geometric in nature, yet we find 
many of its aspects reminiscent of continuous thermal phase transition (CTPT) \cite{ref.Stanley,  ref.Binney}.
Thus, percolation serves as a relatively tractable model for the investigation of phase transition and critical phenomena that lie at the heart of the modern development of statistical
physics. This is perhaps the most important reason why percolation is still studied extensively even after almost 60 years of its inception.

Indeed, for almost every observable quantities in percolation there exist an equivalent counterpart in
CTPT. These observables like their counterpart in CTPT, exhibit power-law, 
at least near $p_c$, which is typically attributed to critical phenomena. For instance, 
the system size $L$ is like correlation length $L\sim (p-p_c)^{-\nu}$, mean cluster size $S$ is like susceptibility $S\sim (p-p_c)^{-\gamma}$, 
percolation strength $P$ is like order parameter $P\sim (p-p_c)^\beta$ etc. 
Like thermal phase transition, percolation transition too can be classified in terms of $p_c$ and by
a set of critical exponents $\beta, \gamma, \nu$ etc. One of the extraordinary findings in percolation is that the numerical value of its critical exponents 
depend neither on the detailed nature of the lattice structure nor on the type of percolation, bond or site. Their values depend
only on the dimension of the embedding space of the lattice.  It is, therefore, said that percolation on all planar lattices belong to the same universality class.

Unique universality class has been found true for a variety of periodic and non-periodic planar lattices having fixed and mixed-valued coordination number, 
random planar lattices and their dual, random multifractal lattices etc. \cite{ref.yonezawa, ref.becker, ref.hsu, ref.multifractal} (see also Ref. \cite{ref.saberi}, which is the most recent review article).
Yet, have we exhausted all the possible lattices to conclude that percolation on all planar lattices belongs to the same universality class? The answer is no. 
Recently, we have reported that the site percolation on a weighted planar stochastic lattice (WPSL) belongs to separate and distinct universality class \cite{ref.hassan_rahman}. The 
WPSL is quite non-trivial as it has mixed properties of both lattice and network or graph \cite{ref.Hassan}. On one hand, unlike networks it is embedded in the space of dimension $d=2$, on the other
 unlike regular lattice, its coordination number distribution obeys a power-law. We found that the critical exponents for site percolation on the WPSL
are totally different from the known values for all other planar lattices studied till to-date. We,
therefore, claim that the random site percolation on the WPSL belong to a separate and distinct universality class.

In this article, we investigate the bond percolation on the WPSL and present 
detailed results of its site counterpart in order to see the contrast.
One of the goals of the present article is to check if the bond and site percolation on WPSL belong to the same universality class like for all known planar lattices studied to date.
First, we find the percolation threshold $p_c$, for both bond and site percolation, using the idea of spanning probability $W(p)$.  
Second, we attempt to find an estimate for the various critical exponents such as $\nu, \beta$ and $\gamma$ using the finite-size scaling hypothesis where precise value of $p_c$ is necessary. Then, we use the idea of 
data collapse for further fine tuning of the estimated values for the exponents till 
we get the best data-collapse. 
Besides critical exponents, we also find the exponent $\tau$ that characterizes the cluster size distribution function $n_s(p_c)\sim s^{-\tau}$ and the fractal dimension $d_f$ that 
characterizes the mass of the spanning cluster $M(p_c)\sim L^{d_f}$. Note that the values of the various critical exponents and the exponents $\tau$, $d_f$ etc. are not at all independent rather they are
bound by some scaling relations. We use these scaling relations for self-consistency check. We find that our estimate for various
exponents satisfy these relations up to quite a good extent. Our results based
on extensive Monte Carlo simulation suggest that both site and bond percolation on WPSL
belong to the same universality class and it is different from the one where percolation on all the planar lattices belong.

The rest of the article is organized as follows. In section II, we discuss the algorithm for the
construction of WPSL and some of its key features. 
In section III, we briefly discuss the Newman-Ziff algorithm as it is the most efficient algorithm
for percolation. We also discuss the finite-size scaling and underline its deep connection
to the Buckingham $\Pi$-theorem in section IV.
In section V, we present our results about bond and site percolation on the WPSL side by side so that
we can appreciate the contrast. Finally, we summarize our
results in section VI.

\section{WPSL and its properties}

\begin{figure}
\includegraphics[width=8.5cm,height=8.0cm,clip=true]{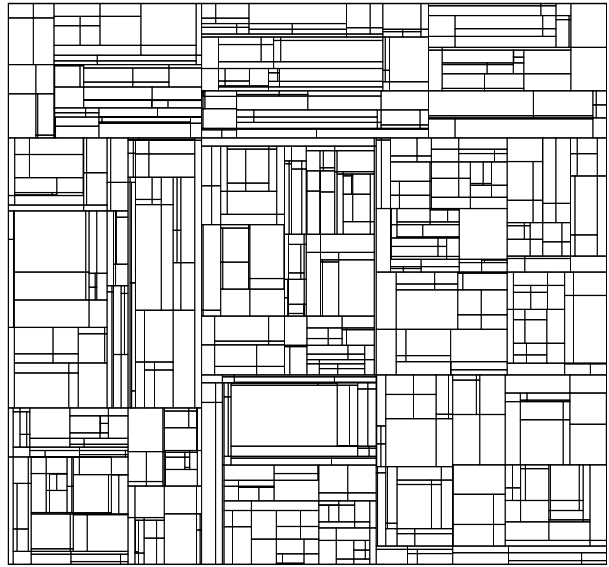}
\caption{A snapshot of the weighted stochastic lattice.
}
\label{fig:1}
\end{figure}%

We first give a brief description of the construction process of the WPSL.
It starts with an initiator which we choose to be a square of unit
area. The generator is then defined as the one that divides the initiator (in step one) randomly into four smaller blocks.
In step two and thereafter the generator is applied to only one of the blocks by picking 
it preferentially with respect to their areas.
Consider the $t$th time step of the generation of the WPSL at which the system has $3t-2$ number 
of blocks available whose areas are say $a_1, a_2, a_3,..., ..., a_{3t-2}$.
To pick one from $3t-2$ blocks we subdivide an interval of unit length $[0,1]$ into $(3t-2)$ 
sub-intervals of size $[0,a_1]$, $[a_1, a_1+a_2]$,$\  ...$, 
$[\sum_{i=1}^{3j-3} a_i,1]$ so that the higher the area the greater the size of the sub-intervals.  
We then generate a random number, say $R$, from the interval $[0,1]$ and 
find which of the $(3t-2)$ sub-intervals contain this $R$ and pick that block. 
This process ensures that the blocks are being picked
preferentially according to their size. In Fig. (\ref{fig:1}) we give a snapshot of the lattice to give a visual
impression of how it actually looks at any given time. It is a space-filling planar cellular structure where the size or the area of the cells in the lattice are not equal 
rather their distribution is random. This is in sharp contrast to many of the cellular structures that we are familiar with. 
One advantage of creating WPSL by random sequential partitioning of the square into ever smaller mutually exclusive rectangular 
blocks helps defining each step of the division process as one time unit. 
The number of blocks $N$ at time $t$ therefore is $N=1+3t$ and hence it 
grows albeit the sum of the areas of all the  blocks is always equal to the size of the initiator. 
Thus, the  number of blocks $N$ increases with time at the expense of the size of the blocks.

Recently, we have shown that the area
size distribution of the blocks of WPSL obey dynamic scaling 
\begin{equation}
c(a,t)\sim t^\theta \phi(a/t^z),
\end{equation}
where we found $\theta=2$ and $z=1$ \cite{ref.hassan_dayeen}. It implies that the snapshots of the lattice at different times are similar.
Yet another interesting properties of this lattice is that the dynamics of the system is governed by infinitely many conservation laws one of which is the conservation of total area. 
To be more precise, if we denote $x_i$ and $y_i$ as the length and width of the $i$th block
then we can show analytically that $M_n=\sum_i^N x_i^{n-1} y_i^{4/n-1}$ assumes statistically 
a constant value regardless
of the time $t$ when the snapshot is taken \cite{ref.Hassan}. We have also shown that, except the conservation of total area,
each of the infinitely conserved quantity is a multifractal measure. 
That is, we can assume that the $i$th block of the lattice is populated with 
probability $p_i\sim x_i^{n-1} y_i^{4/n-1}$. We have shown that within the multifractal 
formalism we can construct the partition function which is 
the $q$th moment of $p_i$ i.e.,
\begin{equation}
\label{eq:1}
Z_q=\sum_i p_i^q.
\end{equation}
Measuring $Z_q$ as a function of the square root of the mean block area 
\begin{equation}
\delta=\sqrt{{{{\rm area\ of \ the \ initiator }}\over{{\rm total \ number \ of \ blocks }}}} =
\sqrt{{{1}\over{1+3t}}} 
\sim t^{-1/2},
\end{equation}
 one can show that $Z_q$ exhibits power-law 
\begin{equation}
\label{eq:2}
 Z_q(\delta)\sim \delta^{-\tau(q,n)},
\end{equation}
with exponent
\begin{equation}
\label{massexponent}
\tau(q,n)=\sqrt{(4/n-n)^2q^2+16}-((4/n+n-2)q+2).
\end{equation} 
One of the characteristic features of this exponent is that it is non-linear $\forall \ n$ except $n=2$.

Note that the exponent $\tau(q,n)$ has two interesting properties. First, $\tau(q,n)=2$ $\forall \ n$
at $q=0$ which is the dimension of the embedding space of the WPSL. Second, 
$\tau(q,n)=0$ $\forall \ n$ at $q=1$ as it is required by the normalization condition \cite{ref.multifractal_1}. 
The Legendre transform of $\tau(q,n)$ is a method whereby its derivative
\begin{equation}
\label{eq:3}
\alpha=-{{d\tau(q,n)}\over{dq}},
\end{equation} 
can be considered as an independent variable instead of $q$ itself. In general, if we denote
$\alpha$ as the slope and $f$ as the intercept then the equation for the straight line is
\begin{equation}
\label{eq:4}
 \tau(q)=-\alpha q+f(\alpha).
\end{equation}
The function $f(\alpha)$ is the Legendre transform of the function $\tau(q)$ which is always concave in character. 
It implies that for every $n$ value there exist a spectrum of spatially intertwined fractal dimensions 
\begin{equation}
\label{eq:5}
f(\alpha(q,n))={{16}\over{\sqrt{({{4}\over{n}}-n)^2q^2+16}}}-2,
\end{equation}
which are needed to characterize the WPSL except for $n=2$. Note that the maximum of $f(\alpha,n)$ occurs at $q=0$ which corresponds to the dimension of the embedding space of the WPSL 
when blocks are assumed empty. We thus find that
the WPSL is a multi-multifractal planar lattice.

Besides, WPSL is a planar cellular structure whose cells or blocks has coordination number disorder in the sense that unlike regular lattice
it has great many different number of neighbors. In fact, its coordination number distribution 
exhibits a power-law \cite{ref.Hassan}. This is in sharp contrast to the coordination number distribution in the Voronoi diagram where it
is also random but its distribution is peaked around the mean \cite{ref.oliveira}. In the
Voronoi diagram it is almost impossible to find 
cells or blocks which have significantly higher or fewer neighbours than the mean coordination number. 
That is, here the mean describes the characteristic scale. Such characteristic scale is absent in the WPSL since the distribution function 
follows a power-law. The power-law coordination number distribution also means that the majority of the 
blocks in the WPSL are very poor in coordination number and there are few cells or blocks which have significantly high number of nearest neighbours. 
A lattice, so rich in properties can be of great interest as it can mimic disordered medium
on which one can study problems like percolation or random walk. 
In brief, the WPSL has the following properties:

\begin{enumerate}
 
 \item[i)] Its area size distribution function obeys dynamic scaling.

 \item[ii)] It obeys infinitely many conservation laws.
 
 \item[iii)] It is a multi-multifractal.
 
 \item[iv)] Its coordination number distribution function obeys power-law.
\end{enumerate}

\section{Newman-Ziff algorithm}

In the standard algorithms, such as the Hoshen-Kopelman (HK), one must create an
entire new state for every given value of occupation probability $p$ in every independent realization.
Investigation of the various observable using such traditional algorithms are highly expensive
in terms of computational time and accuracy of finding various observable quantities. 
In 2000, Newman and Ziff (NZ) proposed an algorithm 
which is highly efficient in both accounts \cite{ref.Ziff}. The efficiency in the NZ algorithm 
lies in the fact that one creates a new state with $n+1$ occupied sites or bonds from
the immediate previous state with $n$ occupied sites or bonds simply by occupying one extra randomly chosen site or bond.
It is based on the intuitive idea of random sequential adsorption of sites or bonds on a given 
lattice or graph. The algorithm is trivially simple. One starts with an empty lattice. Then at each step 
an empty site or bond is chosen at random and then is occupied if empty; else the attempt is discarded. However, in order to further reduce the computation time we first decide an order in which the sites or bonds will be occupied. That is, 
we wish to choose a random permutation of the bonds or sites. This is done by creating 
a list of all the bonds in any convenient order. Positions in this list are numbered from $1,23,...,M$. Choose a number $j$ at random with uniform probability in the
range $i\leq  j\leq M$. Then use any standard textbook algorithm to randomize the number
$i=1$ to $M$ and put them in a new order in which they will be occupied.
Having chosen an order of all the sites, we start occupying them in that order. The first site 
or bond to be occupied will definitely form a cluster of size one. The second, third, fourth etc
too are highly likely to form clusters of size one. However, the likelihood of
forming clusters of size one will decrease with the number of occupied sites since some sites when occupied, will become contiguous occupied sites thus making clusters of size more than one.

The formation of clusters and the statistics of their sizes are the key to the
study of percolation theory. In the case of NZ algorithm 
we measure an observable, say $O$, for fixed numbers of occupied sites (or bonds), 
and obtain a data for $H$ as a function of occupation number $n$.
This is in sharp contrast with the HK algorithm where the number of sites being 
occupied at a given $p$ is random and different at 
every independent realization. However, if the system size is large enough then the mean occupation number 
will almost equal to $pN$ where $N$ represents the system size. The weight factor of obtaining different $n$ for a given $p$ are not the same. The exact weighting factor of there being
exactly $n$ occupied sites on the lattice for a given $p$ is given by binomial distribution
\begin{equation}
C(n,N,p)=\sum_{n=1}^N \left( \begin{array}{c}
N \\ n \end{array}\right )p^n(1-p)^{N-n}.
\end{equation}
The binomial coefficient $\left( \begin{array}{c}
N \\ n \end{array}\right )$ represents the number of
possible configurations of $n$ occupied sites and $N-n$ empty sites. Using this and 
the data for the observable $O$ for all values of $n$ we can find $O$ for any value of $p$ by the following relation
\begin{equation}
\label{eq:convolution}
O(p)=\sum_{n=1}^N \left( \begin{array}{c}
N \\ n \end{array}\right )p^n(1-p)^{N-n} O_n.
\end{equation}
It is interesting to note that the ensemble of states
with exactly $n$ occupied sites or bonds  obtained according to NZ algorithm can referred to
as a {\it microcanonical percolation ensemble}, where the number $n$ is the equivalent counterpart of
the energy $E$ in thermal statistical mechanics. On the other hand, if we keep $p$ fixed instead
of $n$ we can regard it as the canonical ensemble.

\section{Finite-size scaling and $\Pi$-theorem}

We offer here a brief introduction to the spirit and
scope of the scaling approach to phase transitions and
critical phenomena in general. It is well-known as finite-size scaling (FSS) hypothesis. 
It has been extensively used as a very powerful tool for estimating finite size effects
near the threshold value of the controlling parameter.
In the continuous phase transition, the various response functions, typically 
the second derivative of the free-energy, diverges. Such transitions are classified by a set of critical exponents.
The best known example of continuous phase
 transition is the paramagnetic to ferromagnetic
transition where it has has been found that
\begin{eqnarray}
\label{eq:fss_0}
{\rm magnetization}\> \> \> \> & &   M\sim (T-T_c)^\beta, \nonumber \\
{\rm susceptibility} \>\>\> \> & & \chi_M\sim (T-T_c)^{-\gamma}, \nonumber \\
{\rm and \ correlation \ length} \>\>\> \> & &  \xi \sim (T-T_c)^{-\nu}.
\end{eqnarray}
In percolation, their equivalent counterparts are
\begin{eqnarray}
\label{eq:fss_1}
{\rm percolation \ probability}\> \> \> \> & &   P \sim (p-p_c)^\beta, \nonumber \\
{\rm Mean \ cluster \ size} \>\>\> \> & & S\sim (p-p_c)^{-\gamma}, \nonumber \\
{\rm and \ system \ length} \>\>\> \> & &  \xi \sim (p-p_c)^{-\nu}.
\end{eqnarray}
These relations are only true in the thermodynamic limit in the sense that the system size is infinite. It is important to appreciate the fact that we can neither do experiment nor simulation on infinite systems where the correlation length 
$\xi\sim L$. To overcome this impediment, physicists have come up with a smart solution which
is known as finite-size scaling. In general, an observable quantity, say $X$, of the threshold phenomena that exhibit continuous phase transition is said to obey finite-size scaling if it satisfies
\begin{equation}
X(p,L)\sim L^{a/\nu}\phi((p-p_c)L^{1/\nu}),
\end{equation}
where $a$ and $\nu$ are said to be critical exponents.
It provides an elegant way of extrapolating critical exponents for infinite system from a set of 
data for finite systems using the idea of  data collapse.

We shall here show that the origin of the FSS theory is actually deeply rooted to the Buckingham $\Pi$-theorem
as it can be systematically obtained following the prescription of that theorem \cite{ref.barenblatt}. 
Consider that a quantity $X$ is the
primary quantity of interest which depends on the control parameter $x$ and the system size $L$ so that
we can write
\begin{equation}
X=X(x,L).
\end{equation}
Note that in the case of threshold phenomena, where there is a critical or threshold value $x_c$ across which
the system under goes a sudden or abrupt change, we find that the distance $x-x_c$ is a better variable than $x$ itself. Indeed, the observable quantity $X$ is found to depend on $x-x_c$ and hence we write 
\begin{equation}
\label{eq:fss_2}
X\sim X(x-x_c,L).
\end{equation}
We almost always find that the quantity $x-x_c$ diminishes with $L$ following a power-law 
$(x-x_c)\sim L^{-a}$. It implies that we can 
choose one of the parameters, say $L$, to have an independent dimension. Thus the dimension of $X$ too
can be expressed in terms of $L$ alone
\begin{equation}
X\sim L^b.
\end{equation}
Following the argument of the $\Pi$-theorem we can now define two dimensionless quantities
\begin{equation}
\xi={{x-x_c}\over{L^{-a}}},
\end{equation}
and
\begin{equation}
\Pi={{X}\over{L^b}}\equiv \phi(\xi,L).
\end{equation}
Note that $\phi$ being a dimensionless quantity its numerical value must remain invariant,
for a given value of $\xi$, even if we change $L$ by an arbitrary factor and hence 
$\phi(\xi,L)=\phi(\xi)$. We can thus immediately write that
\begin{equation}
\label{eq:fss_4}
X(x,L)\sim L^{b}\phi((x-x_c)/L^{-a}).
\end{equation}
The reduction of initially two variable problem into one variable problem constitutes the basic
statement of the Buckingham $\Pi$-theorem. This is traditionally 
known as an hypothesis in the literature namely as the finite-size scaling hypothesis.

A quantitative way of interpreting how the experimental data exhibit finite-size scaling is done by 
invoking the idea of the data-collapse method  - an 
idea that goes back to the original observation of Rushbrooke \cite{ref.Stanley}. The plots of $X(x,L)$ vs $x$ for
different $L$ always result in distinct curves. However, the same data can be made to collapse 
on a single universal curve if one plot $XL^{-b}$ vs $(x-x_c)L^{a}$ instead of $X(x,L)$ vs $x$
regardless of the size of $L$. The quality of data collapse depends on how exact the value of $x_c$
and the exponents $a$ and $b$. Data-collapse 
means that the characteristic properties of the system represented by $X$ are similar on different system
size $L$. Note that two systems of different sizes are said to be similar if they differ in the numerical 
value of their dimensional quantities $X$ and $x$, however, the numerical value of the corresponding dimensionless quantities 
$XL^{-b}$ and $(x-x_c)L^{a}$ coincide and that is why we obtain data-collapse.
Obtaining data-collapse guarantees that the system exhibits scaling or similarity with respect
to different independent system size. It is an extension of the idea of similarity of two triangles.
For instance, two right triangles (characterized by their area $S$ and the sides $a, b$ and the
hypotenuse $c$) may differ in the 
numerical value of their dimensional quantities. Now, one can vary $b$ keeping $a$ fixed and measure $S$
for both the triangles. Plotting $S$ as a function of $b$ will definitely give two distinct curves one for
each. However, the plots of the corresponding dimensionless quantities $S/c^2$ vs $b/c$ will give rise to single universal curve since the numerical value of $S/c^2$ will 
always coincide for a given value of the 
acute angle $\theta$ regardless of the size of the triangle. This happens because triangles are similar.

\section{Site/bond percolation on WPSL}

What is site and bond in WPSL? Before answering this question we find it worth 
discussing first what they are in the context of conventional lattices. For instance, we can regard a square 
lattice as a grid or mesh. Each cell of the grid has four sides and each side is
a common border of two cells only. In the case of square grid, we can thus regard
each cell as a site since it contains exactly one lattice point. Equivalently, 
we could also regard the vertices of each cell as sites. However, in the present context we 
stick to the former definition. The dual of the square grid, obtained by replacing the center 
of each cell by a node and the common border between neighbouring cells by a link connecting the two nodes. 
We can thus regard the links
of the dual as the bond of the square lattice. Following the same argument we regard 
the blocks of the WPSL as its sites not the vertices of the lines that tessellated the initiator. 
To define bond, we first find its dual. 
It is obtained by replacing the center of each block by a node and the common border 
between two neighbouring blocks by a link connecting the corresponding nodes. We regard these links 
as the bonds of the WPSL. Using these ideas we first performed site and bond percolation 
on the square lattice and reproduced all the known results and then we applied them to the WPSL. 

Recently, we have studied site percolation on WPSL, and found non-trivial
results. That is, it belongs to a separate universality 
class than the universality class where percolation on all planar lattices are believed to belong. 
However, we are yet to check whether the site
and bond percolation on WPSL belong to the same class or not.
The dual of the WPSL can be well described as complex network and we have shown in Ref. \cite{ref.Hassan} that
the corresponding degree distribution follow a power-law $P(k)\sim k^{-\gamma}$ with exponent 
$\gamma=5.58$. Interestingly, the degree distribution $P(k)$ in the context of network is the same as
the coordination number distribution in the context of lattice. However, there is
a sharp difference between networks based on graph theory and the network obtained
from the dual of a lattice which is embedded in a space. 
The difference lies in the fact that networks based on graph theory 
have no edge or surface but networks based on the dual of a lattice have edge or surface which is
crucial in the case of percolation as it is useful in defining the spanning cluster.

In the case of bond percolation, the lattice consists initially of $N$ blocks and hence
the system has exactly $N$ number of cluster of size one since the center of each block represents a site.
Thereafter, each time we occupy a bond, a cluster at least of size two or more is formed. 
In the case of site percolation, each time we occupy a block, the size of the cluster may vary 
as we measure it by the area of contiguous occupied blocks. Initially all the blocks are empty
and we won't know the size of the cluster even after the first block is occupied. For regular
lattice like square lattice of $L^2$ sites have $2L(L-1)$ and $2L^2$ bonds with open and 
periodic boundary condition respectively. Now in the case of WPSL, being a disordered lattice, we cannot have
such exact relation. We still find that the number of bonds or sites when
we take average over ensemble of independent realizations follow a relation valid for all size of the
lattice. For instance, for the lattice at time $t$ there are exactly $3t+1$ sites and on the average there
are $8t$ bonds with periodic boundary condition. Thus the mean coordination number is equal to $16t/3t\sim 5.33$ which is higher
than the square lattice. We know that the percolation threshold $p_c$ depends on coordination number of the lattice and
the higher the mean coordination number of a lattice the lesser is the value of $p_c$. In the case of 
for square lattice, for instance, each site has exactly four nearest neighbours and each bond has six and hence $p_c$ of site
percolation is higher than that of the bond. In the case of WPSL, we find that the mean number
of nearest neighbors of a bond is $10.01$ which is almost double the mean nearest neighbour of a site. 
So, it is expected that the $p_c$ value for bond
percolation in WPSL will be quite less than $p_c=0.5265$ for the site percolation \cite{ref.hassan_rahman}.

Percolation is all about formation of clusters and the statistics of their various
properties as a function of control  parameter
$p$ and $L$. The typical observable quantities in percolation are (i) Spanning probability $W(p)$,
(ii) percolation probability or percolation strength $P$,
(iii) The mean cluster size $S$, (iv) cluster size distribution function $n_s(p)$ etc and their variation with $p$ or $L$.

\subsection{Spanning probability $W(p)$}

\begin{figure}
\centering
\subfloat[]
{
\includegraphics[height=4.0 cm, width=2.4 cm, clip=true,angle=-90]
{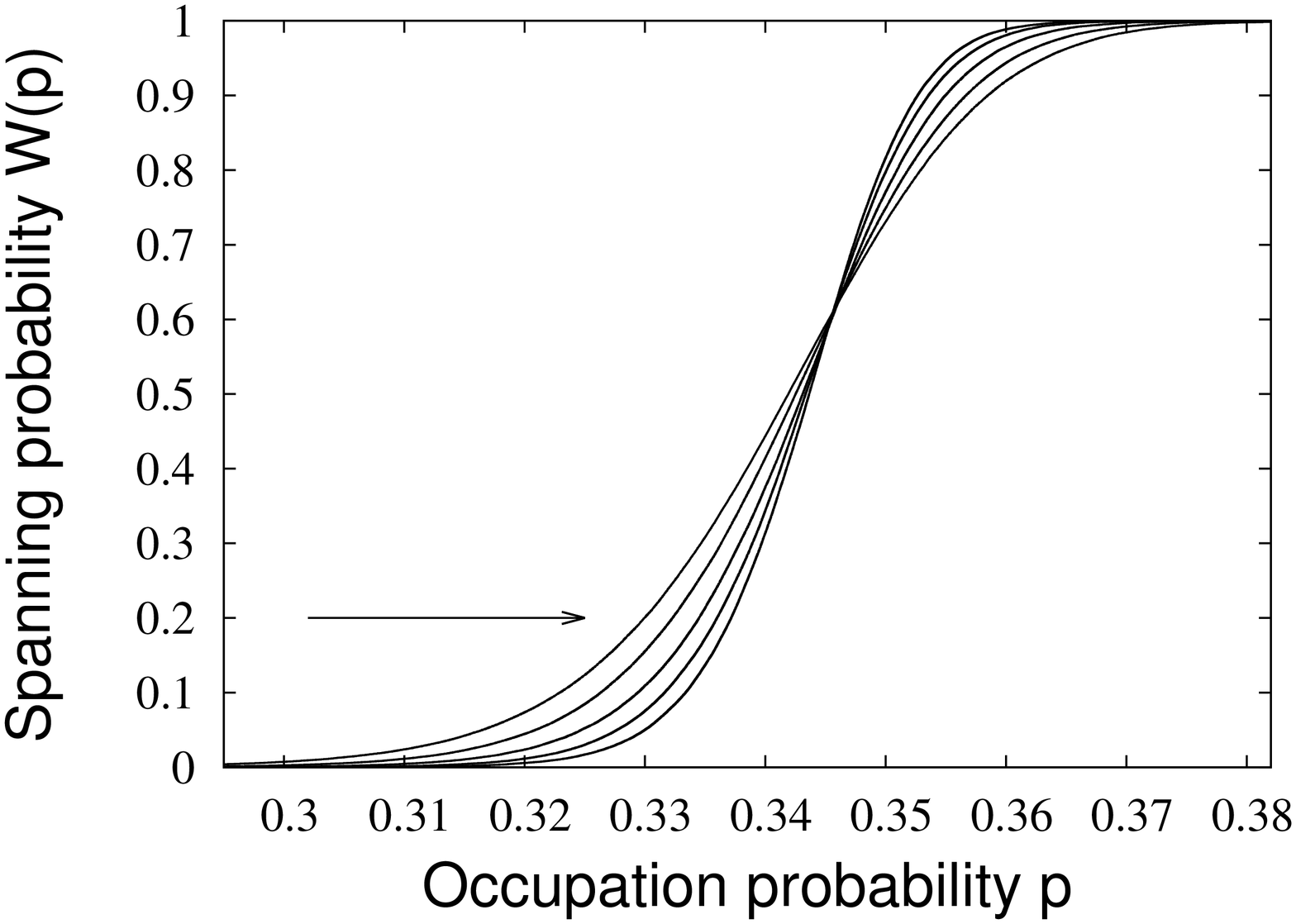}
\label{fig:2a}
}
\subfloat[]
{
\includegraphics[height=4.0 cm, width=2.4 cm, clip=true, angle=-90]
{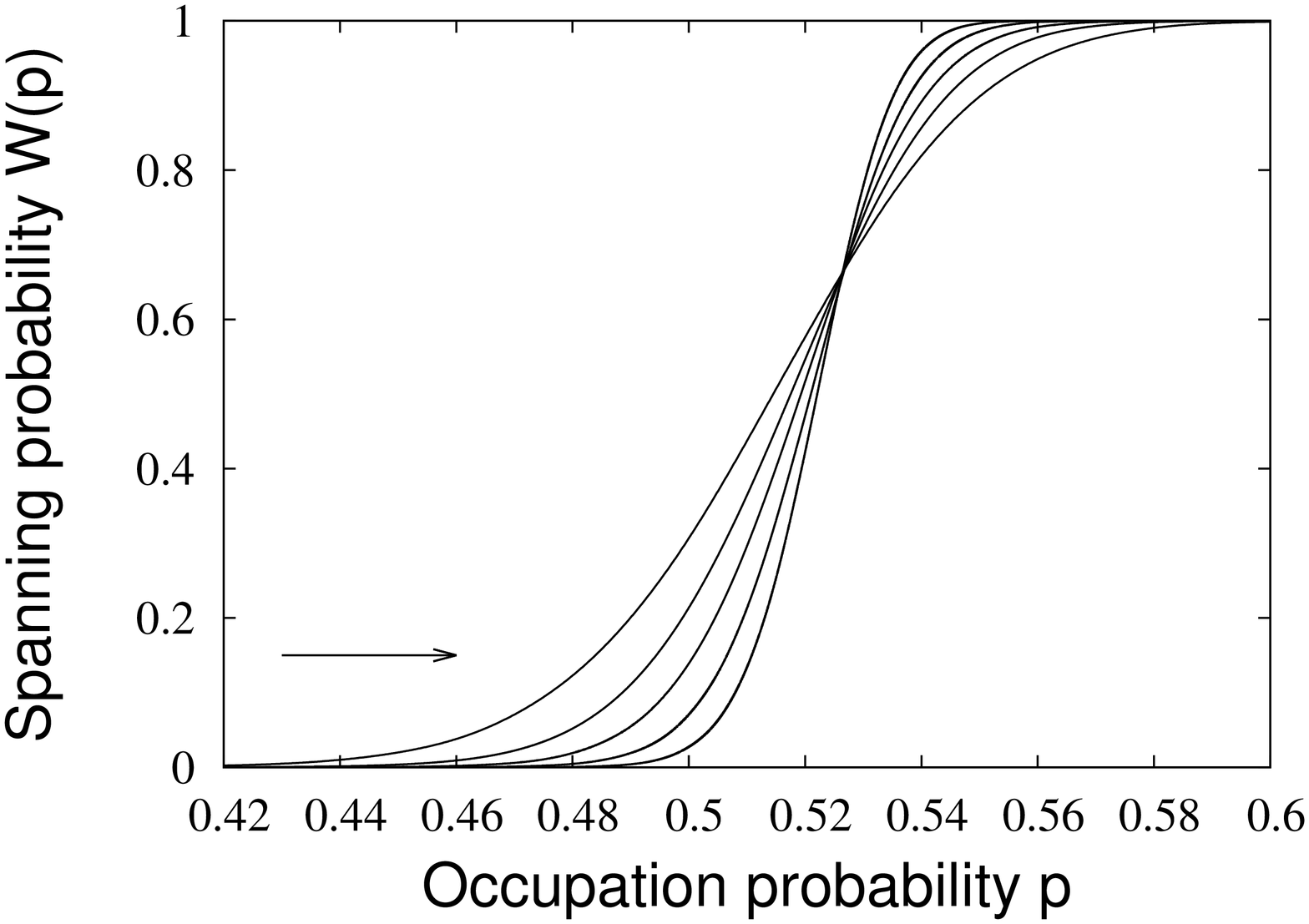}
\label{fig:2b}
}

\subfloat[]
{
\includegraphics[height=4.0 cm, width=2.4 cm, clip=true, angle=-90]
{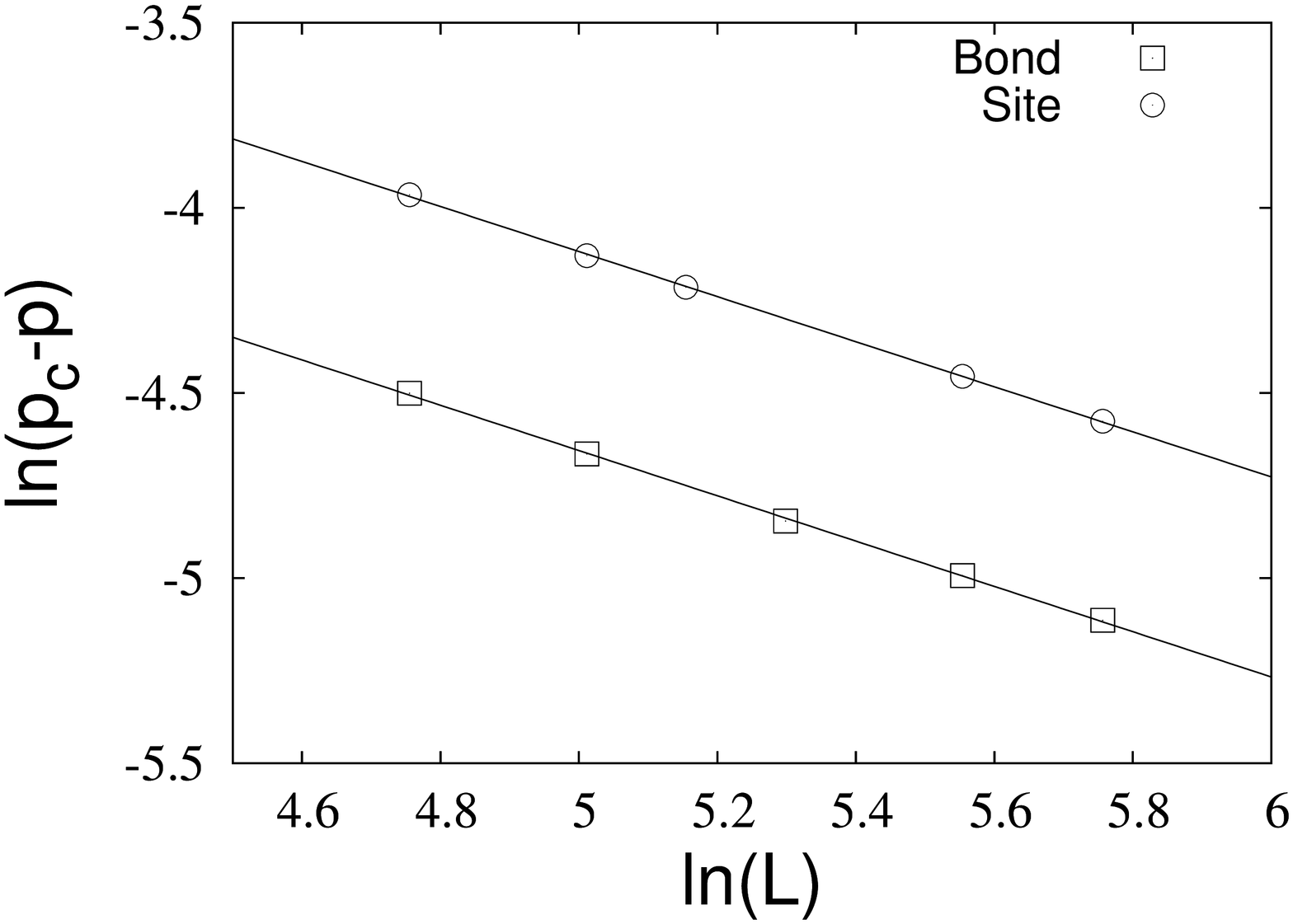}
\label{fig:2c}
}
\subfloat[]
{
\includegraphics[height=4.0 cm, width=2.4 cm, clip=true, angle=-90]
{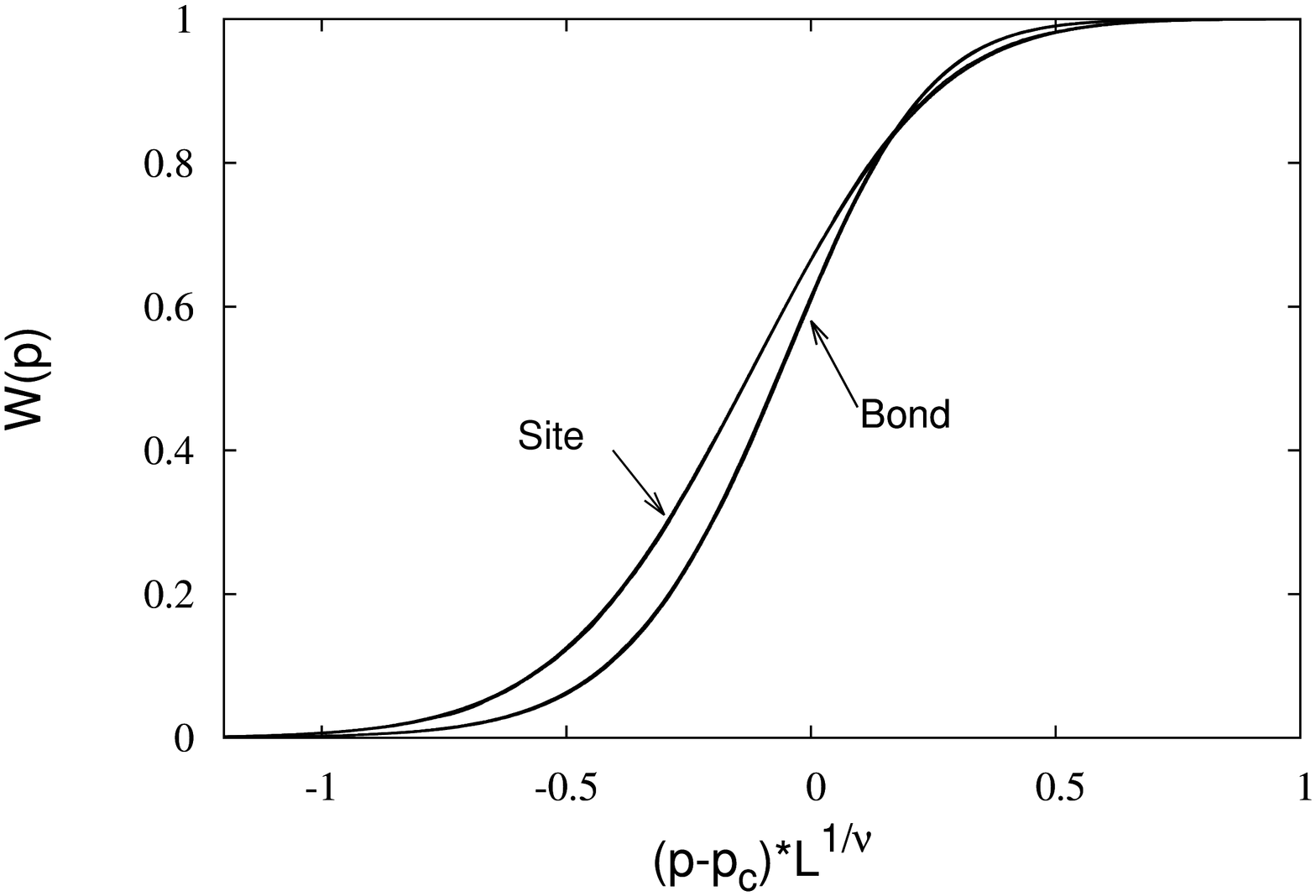}
\label{fig:2d}
}

\caption{Spanning probability $W(p,L)$ vs $p$ in WPSL for (a) bond and 
(b) site percolation. The simulation result of the percolation threshold is
$p_{c}=0.3457$ for bond and $0.5265$ for site. In (c) we plot $\log(p-p_c)$ vs $\log L$ for both bond and site. The two
lines have slopes $1/\nu=0.611714\pm 0.007459$ and $0.613552 \pm 0.003861$ for bond and site respectively. 
In (d) we plot dimensionless quantities $W$ vs $(p-p_c)L^{1/\nu}$ and by tuning the
$\nu$ value slightly we find an excellent data-collapse using $1/\nu=0.6115$ in both the cases
which implies that the $1/\nu$ is aproximately independent of the type of percolation. 
} 
\label{fig:2abcd}
\end{figure}

The spanning probability $W(p)$ for both bond and site describes the likelihood of finding a 
cluster that spans across the system
either horizontally or vertically at the occupation probability $p$. To find how $W(p)$
behaves with the control parameter $p$ we perform many, say $M$, independent realizations 
under the same identical conditions. In each realization for a given finite system size we take record of the 
$p_c$ value at which the spanning cluster appears for the first time. To find a regularity or a pattern
among all the $M$ numbers of $p_c$ values recorded, one usually looks at the relative frequency of occurrence within a class or width $\Delta p$. 
To find $W(p)$, we can process the data containing $M$ number of $p_c$ values to plot histogram displaying normalized relative frequency as a function of class of 
width $\Delta p$ chosen as per convenience. 
In Figs. (\ref{fig:2a}) and (\ref{fig:2b}) we show a set of plots of $W(p)$ for bond and site percolation respectively as a function of
$p$ where distinct curves are for different system size $L=\sqrt{N}$. One of the significant features of such plots is that they all 
meet at one particular $p$ value regardless of the value of $L$. 
It means that even if we had data for infinite system the resulting plot would still meet at the same point
revealing that it must have a special significance and the significance is that it is the threshold  probability $p_c$. Note that finding the $p_c$ value for different
lattice is one of the central problems in percolation theory. In the case of bond we find $p_c=0.3457$ which
is exceedingly less than its site counterpart since on the average nearest neighbor
that each bond has in the WPSL is much higher than for its site counterpart.

The second most significant feature of the $W(p)$ vs $p$ plot is the direction of shift of the curve 
on either side of $p_c$ as the system size $L$ increases. 
This shift with $L$ clearly reveals that all the data points, i.e. the $p$ values, 
are marching towards $p_c$. We can quantify the extent at which they are marching by
measuring the magnitude of the difference $(p_c- p)$ for different $L$. That is, we
can draw a horizontal line at a given value of $W$, preferably at the position where this difference is
the most, and take
records of the difference $p_c-p$ as a function of system size $L$.
Plotting the resulting data after taking log of both the variables or in the logarithmic scale we
find a straight line whose slope gives an estimate of the inverse of $1/\nu=0.613552 \pm 0.003861$
since Fig. (\ref{fig:2c}) suggests
\begin{equation}
\label{eq:sp1}
p_c- p\sim L^{-{{1}\over{\nu}}}.
\end{equation}
It implies that in the limit $L\rightarrow \infty$ all the $p$ takes the
value $p_c$ revealing that $W(p)$ will ultimately become a step function so that $W(p)=0$ 
for $p\leq p_c$ and $W(p)=1$ for $p>p_c$. We can use Eq. (\ref{eq:sp1}) to define a dimensionless quantity $(p_c- p)L^{{{1}\over{\nu}}}$.
Now,  we plot $W(p)$ vs $(p_c- p)L^{{{1}\over{\nu}}}$ in Fig. (\ref{fig:2d}) and 
we see that all the distinct plots  $W(p)$ vs $p$ for bond percolation collapse onto a one universal curve and for site onto another curve albeit they share the same
$\nu$ value. By tuning the $1/\nu$ value further we can get an excellent data-collapse for $1/\nu=0.6115$
and hence a better
$\nu\sim 1.635$ value that corresponds to infinite lattice size.

\subsection{Percolation probability $P$}

Consider that we pick a site at random and ask: How likely is that site belong to the spanning cluster?
For finite system size, it may not belong to the spanning cluster even if $p$ is larger than 
the percolation threshold $p_c$.
Therefore, we therefore can quantify the strength of the spanning cluster by percolation probability $P$ which describes how likely 
a site picked at random is to belong to the spanning cluster. The quantity $P$ is defined as the
ratio of the size of the spanning cluster $s_\infty$ to the size of the lattice $N$ i.e.,
\begin{equation}
\label{eq:pp1}
P={{{\rm Number \ of \ sites \ in \ the \ spanning \ cluster}}\over{{\rm Total \ 
number \ of \ sites \ in \ the \ lattice}}}.
\end{equation}
Sometimes, percolation probability is also defined
as the probability that an occupied site belongs to the spanning cluster. It can be obtained if
we replace the denominator $N$ of Eq. (\ref{eq:pp1}) by total occupied sites. We, however, will
consider the former definition. There exists yet another definition where we can use the size of the largest cluster instead of the spanning cluster. Note that all of these definitions behaves in the same fashion
like order parameter. That is,  in the limit $L\rightarrow \infty$, $P=0$ for $p\leq p_c$ and it rises from $P=0$ at $p_c$ to $P=1$ continuously and monotonically like $P\sim (p-p_c)^\beta$. 
Such behavior is reminiscent of order parameter like magnetization $m$ in the case of paramagnetic to ferromagnetic transition and
hence $P$ is regarded as the order parameter in percolation theory. The critical exponent
$\beta$ value is known to depend only on the dimension of the lattice and independent of the type of percolation. Through the site
percolation on WPSL we already reported that $\beta$ value for WPSL, which is a planar lattice, is different from the value for all the known lattices whose dimension of the embedding space $d=2$.  
We shall now check if the $\beta$ value for the bond percolation is the same as for the site percolation.

\begin{figure}
\centering
\subfloat[]
{
\includegraphics[height=4.0 cm, width=2.4 cm, clip=true,angle=-90]
{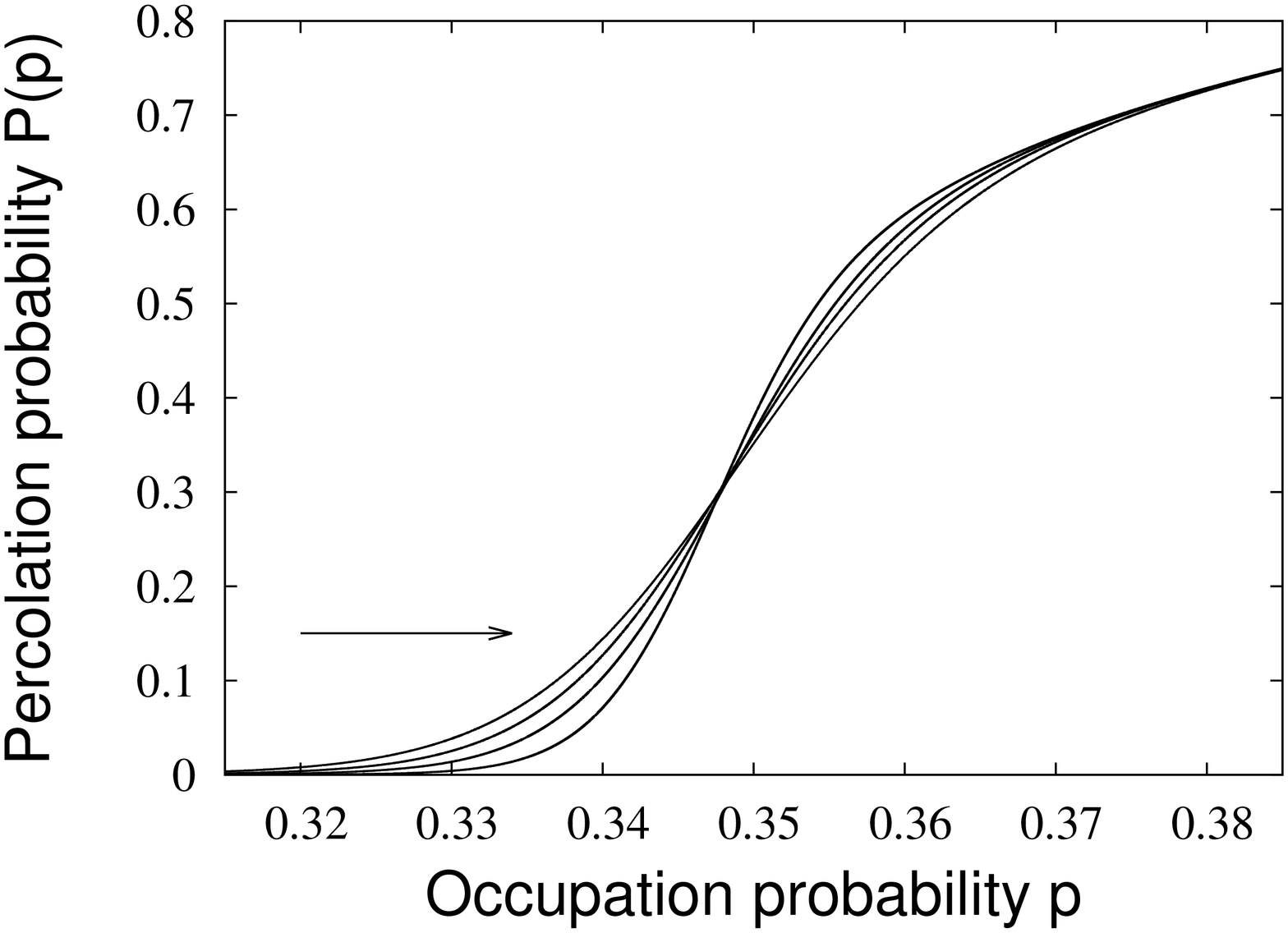}
\label{fig:3a}
}
\subfloat[]
{
\includegraphics[height=4.0 cm, width=2.4 cm, clip=true, angle=-90]
{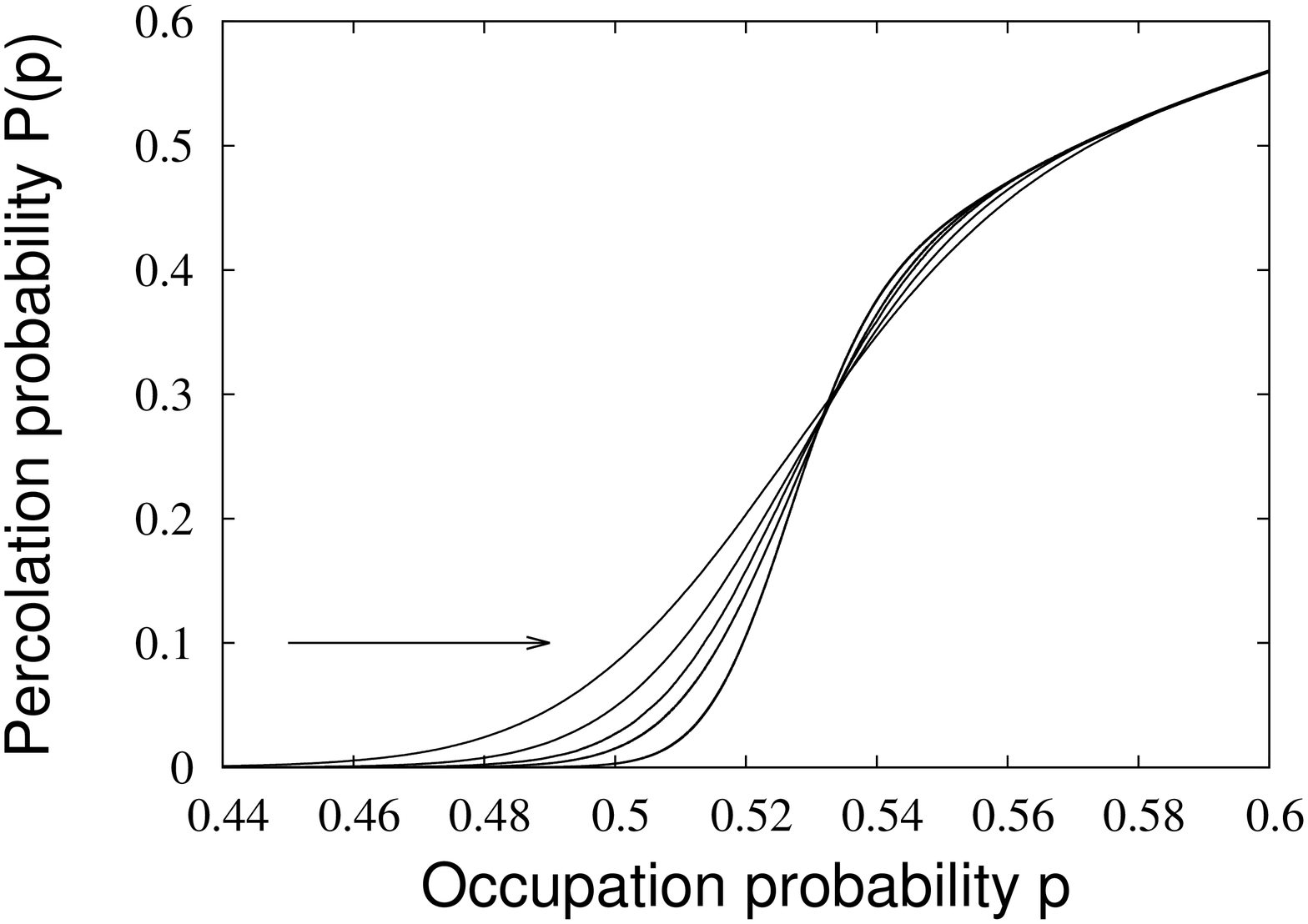}
\label{fig:3b}
}

\subfloat[]
{
\includegraphics[height=4.0 cm, width=2.4 cm, clip=true, angle=-90]
{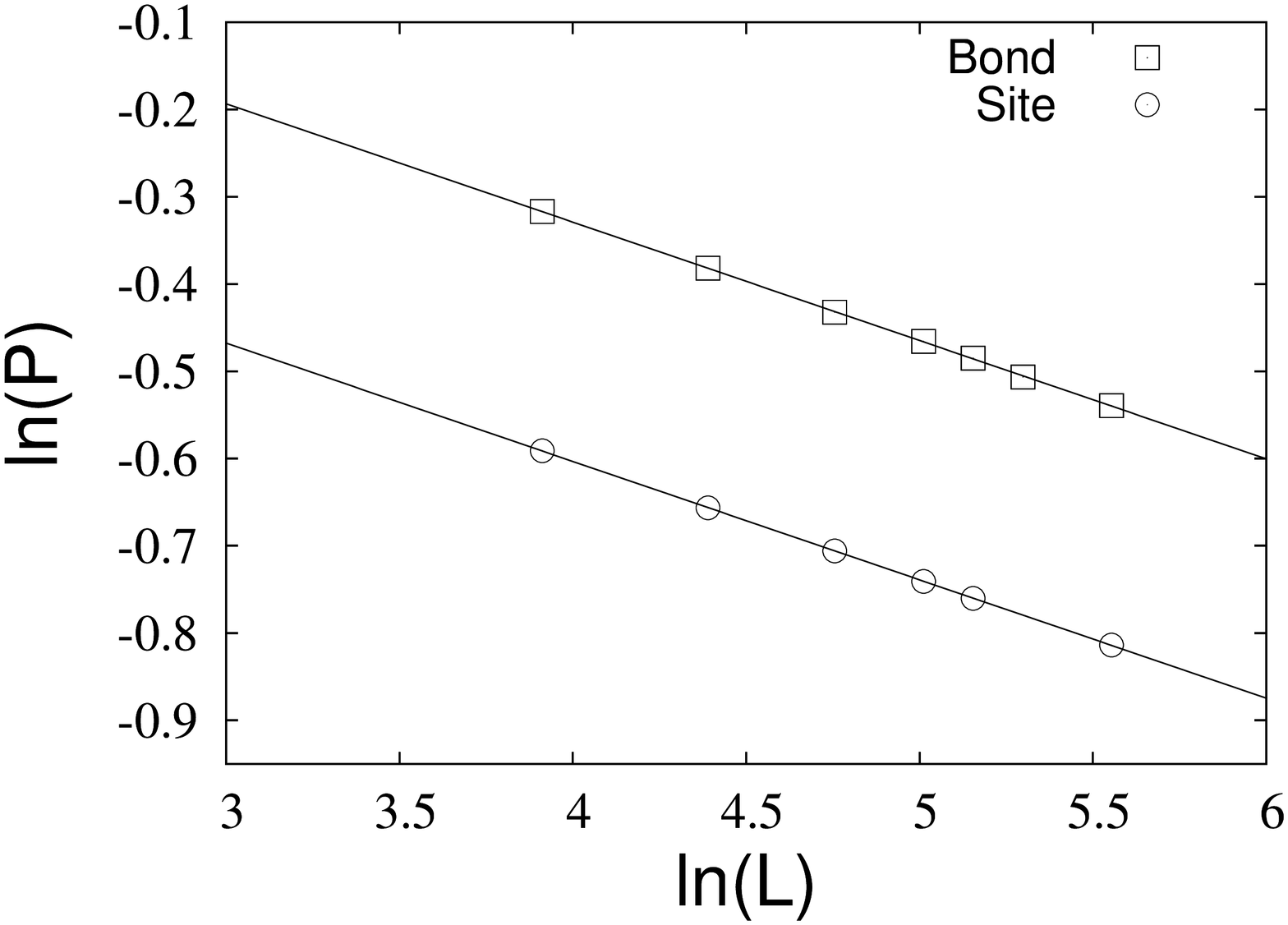}
\label{fig:3c}
}
\subfloat[]
{
\includegraphics[height=4.0 cm, width=2.4 cm, clip=true, angle=-90]
{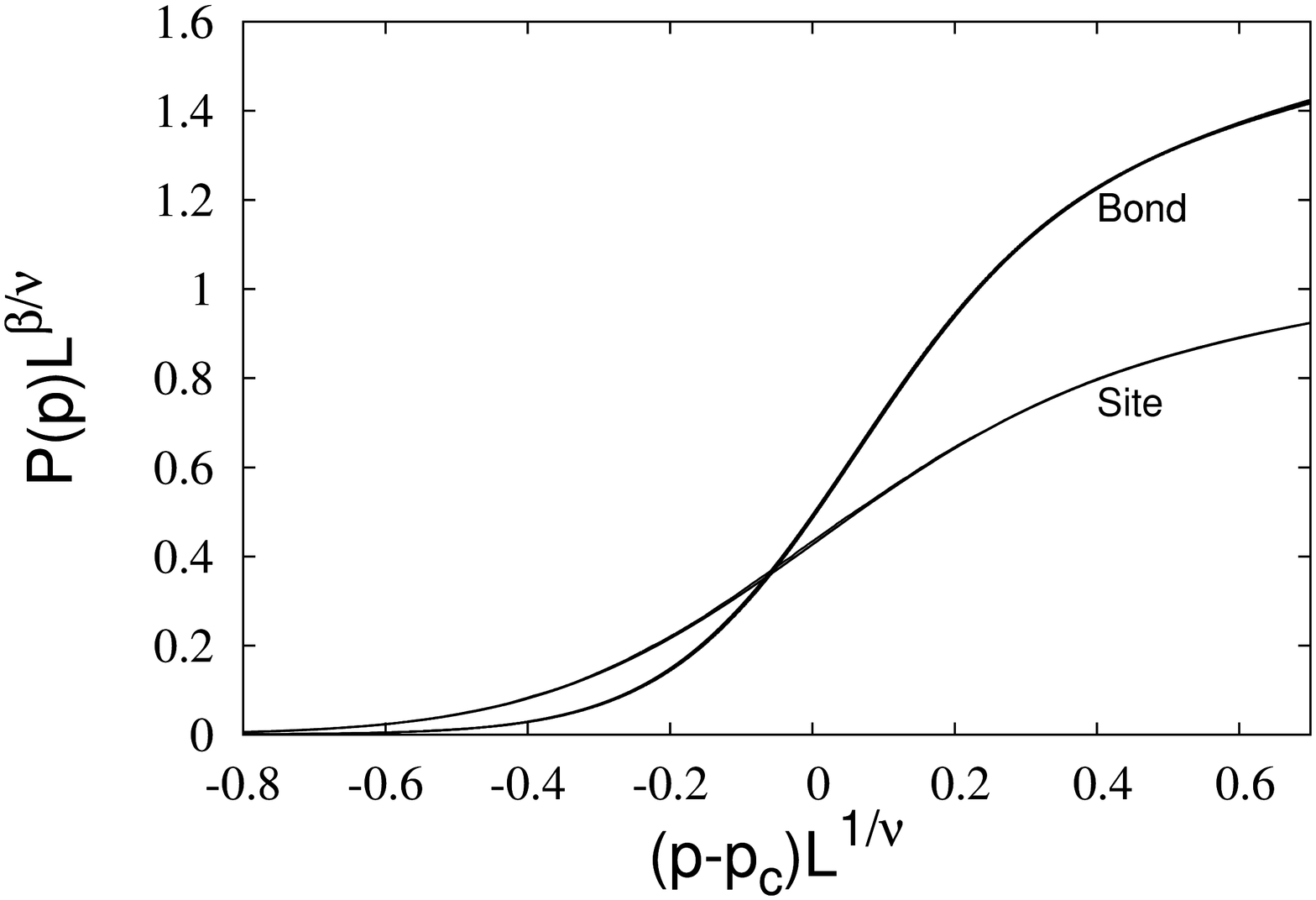}
\label{fig:3d}
}

\caption{Percolation strength or percolation probability $P(p,L)$ in WPSL for (a) bond and (b) site percolation.
In (c) we plot $\log P$ vs $\log L$ using data for fixed value of $(p-p_c)L^{1/\nu}$ and find almost
parallel lines with slopes $\beta/\nu=0.135699 \pm 0.0005905$ for bond and $ 0.135701 \pm 0.0002768$ for site respectively which clearly implies that the critical exponent $\beta$ 
is independent of the type of percolation. For further fine tuning of the $\beta$ value 
we also plot the same data of (a) and (b) in the self-similar coordinates namely
$PL^{\beta/\nu}$ and $(p-p_c)L^{1/\nu}$ and find excellent data-collapse of the plots (a) and (b) both using $\beta/\nu=0.1357$ which gives $\beta\sim 0.222$
} 
\label{fig:3abcd}
\end{figure}

It is important to note that in the case of site percolation we occupy its blocks or cells which
are of different size. We therefore measure the area of the spanning cluster, not the number of 
blocks in the spanning cluster. This is in sharp contrast to the regular lattice where all the blocks or cells are of the same size and hence the size of the 
spanning clusters can be described by the number of blocks or sites in the spanning cluster. In the case of bond percolation on WPSL we, however, use the traditional definition of cluster size. 
This is one significant difference between bond and site percolation on WPSL. Note that for bond 
percolation on WPSL we use the dual of the WPSL not the lattice itself. The dual of the WPSL is obtained
by replacing each block of the WPSL by a node or vertex at its center and each common border between blocks by a bond connecting the nodes at the center of corresponding blocks. In the case of bond percolation 
we occupy these links and measure the size of the cluster by the number of nodes or vertices that the
cluster contains. Below we shall see the impact of this difference in their behavior, if at all. 
In Figs. (\ref{fig:3a}) and (\ref{fig:3b}) we plot percolation probability $P$ as a function of $p$ for bond and site respectively. Looking at the plots, one
may think that all the plots for different $L$ meet at a single unique point like it does for $W(p)$ vs $p$ plot. However, if one zoom in it becomes apparant it is not so and hence the
$p_c$ value from this plot will not be as satisfactory as it is from $W(p)$ vs $p$ plot. 
We also find that $P(p)$ is not 
strictly equal to zero at $p<p_c$, rather there is always a non-zero chance of finding a spanning cluster
even at $p<p_c$ as long as the system size $L$ is finite. However, the plots of $P$ vs $p$ for different system size $L$ reveals that the chances of getting 
spanning cluster at $p<p_c$ diminishes with increasing $L$. There is also a lateral shift of $P$ value
to the left for $p>p_c$ but the extent of this shift $p-p_c$ decreases to such an extent that it never diminishes.
On the other hand, the extent of shift $p-p_c$ to the right for $p<p_c$ diminishes to zero
following Eq (\ref{eq:sp1}).  We shall now check if $P$ above $p_c$ grows like $P\sim (p-p_c)^\beta$. If it does 
so then we shall find the value of the critical exponent $\beta$ and compare it with that of its site counterpart. 

To show that the percolation probability behaves like $P\sim (p-p_c)^\beta$  
and to find the exponent $\beta$ for infinite system size $L$ we use the idea of finite-size scaling. 
We first plot $P(p)$ vs $(p_c- p_c(L))L^{{{1}\over{\nu}}}$
and find that unlike $W(p)$ vs $(p_c- p_c(L))L^{{{1}\over{\nu}}}$ it does not collapse. 
Instead, we find that for a given value of $(p-p_c)L^{1/\nu}$ the $P$ value 
decreases with lattice size $L$. It means percolation probability is not a dimensionless quantity
and hence assume that 
\begin{equation}
\label{eq:pp2}
P\sim L^{-a},
\end{equation}
and we choose $a=\beta/\nu$ for later convenience. 
To find the value of $\beta/\nu$ we measure the heights at a given value of $(p-p_c)L^{1/\nu}$
for different $L$ and plot them in the log-log scale. We find straight lines for both bond and site (see Fig. (\ref{fig:3c})) with slopes $\beta/\nu=0.135699 \pm 0.0005905$ for bond
and $ 0.135701 \pm 0.0002768$ for site revealing that they are almost parallel.
It implies that if we now plot $PL^{\beta/\nu}$ vs $(p-p_c)L^{1/\nu}$ all the distinct plots of $P$ vs $p$ should collapse into a single universal curve. In Fig. (\ref{fig:3d}) we plot just that and 
find an excellent data-collapse using $\beta/\nu=0.1357$ for both bond and site. We checked it for square lattice anyway. 
This again implies that percolation probability $P$
exhibits finite-size scaling
\begin{equation}
\label{eq:pp3}
P(p_c-p,L)\sim L^{-\beta/\nu}\phi\Big ((p-p_c)L^{1/\nu}\Big ).
\end{equation}
Note that although the critical exponents of both site and bond coincide their collapsed universal 
curve does not. We have chacked it with the site and bond percolation on square lattice and found that there too the universal curve do not coincide. Hsu and Huang also stated that the universal curves are different
for planar random lattice, dual of the planar random latice and of the square lattice albeit they belong
to the same universality class \cite{ref.hsu}. 
Now using Eq. (\ref{eq:pp2}) in Eq. (\ref{eq:pp3}) to eliminate $L$ in favor of $p-p_c$  we get
\begin{equation}
\label{eq:pp4}
P\sim (p-p_c)^\beta,
\end{equation}
where $\beta\sim 0.222$ independent of site or bond percolation and it is significantly different from the corresponding values for all known planar lattices.

\subsection{Cluster size distribution and their mean}

\begin{figure}
\centering
\subfloat[]
{
\includegraphics[height=4.0 cm, width=2.4 cm, clip=true,angle=-90]
{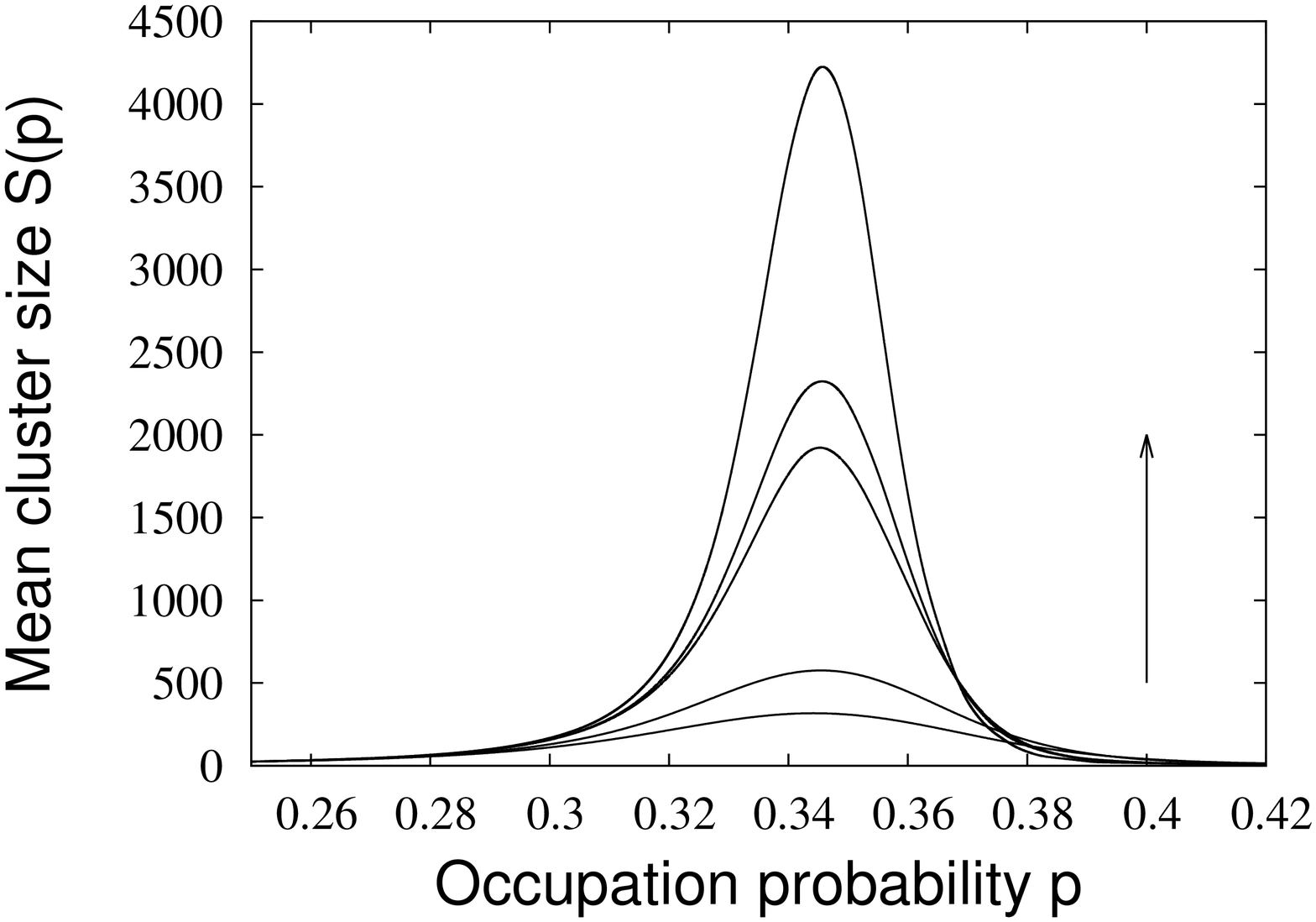}
\label{fig:4a}
}
\subfloat[]
{
\includegraphics[height=4.0 cm, width=2.4 cm, clip=true, angle=-90]
{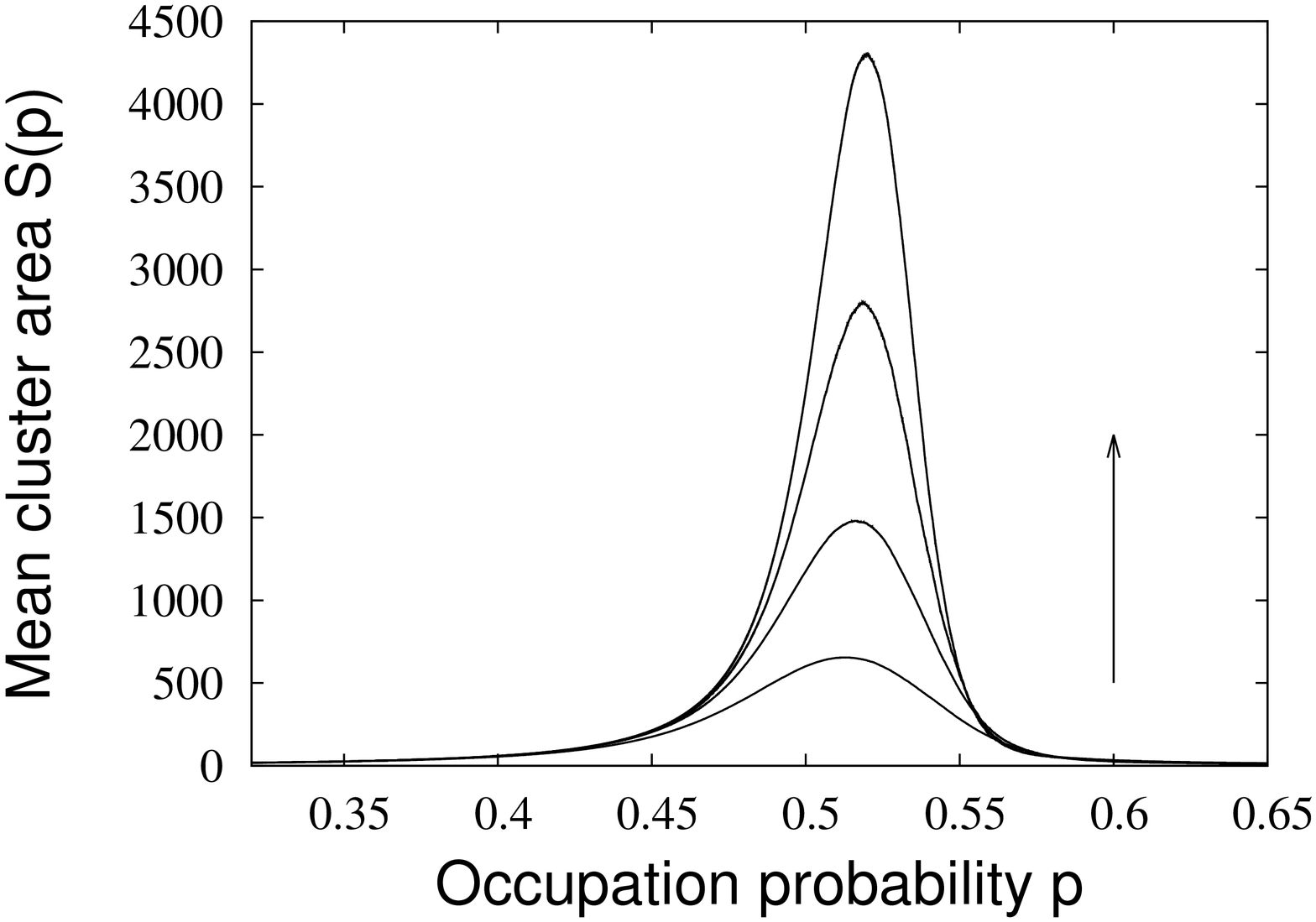}
\label{fig:4b}
}

\subfloat[]
{
\includegraphics[height=4.0 cm, width=2.4 cm, clip=true, angle=-90]
{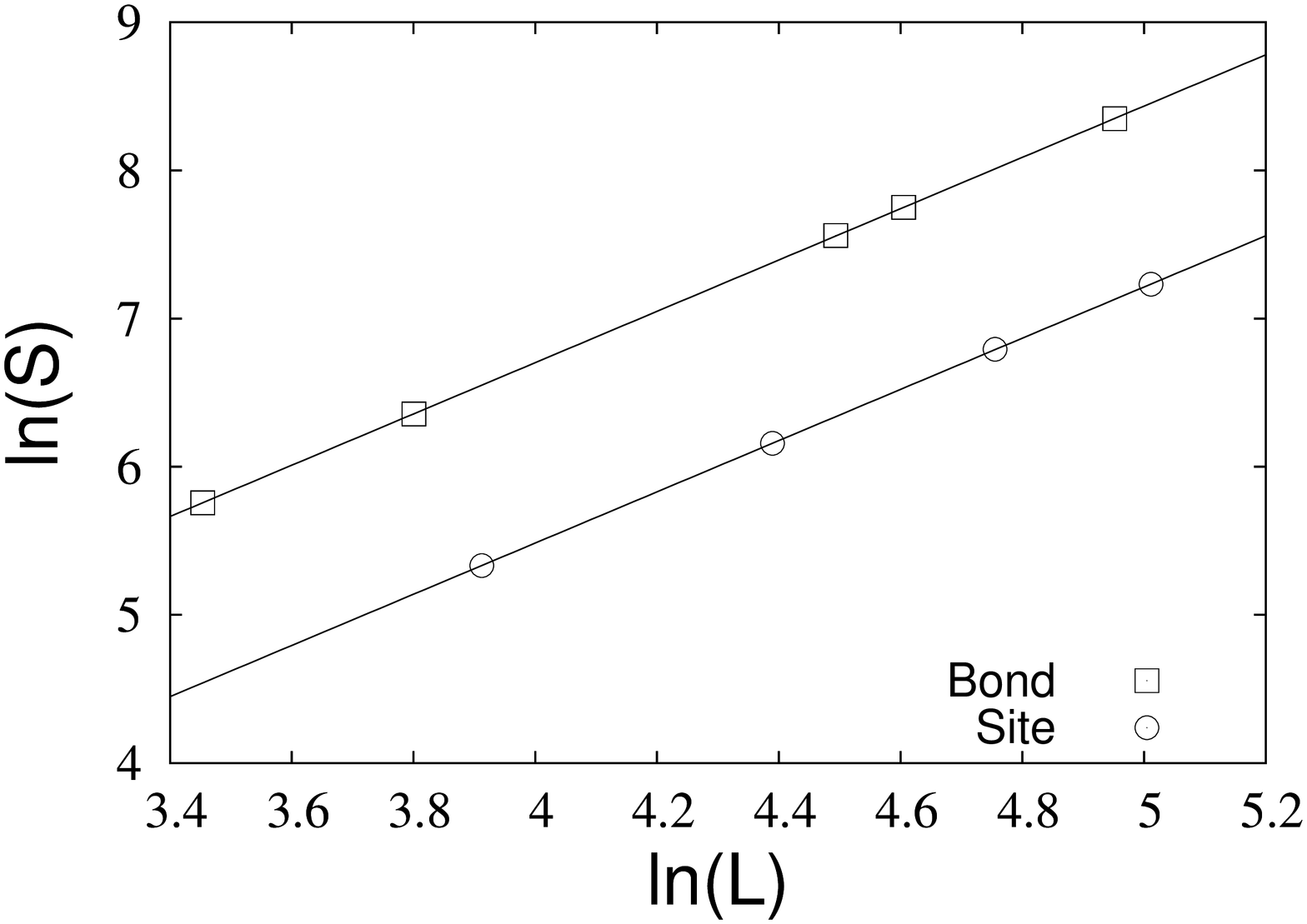}
\label{fig:4c}
}
\subfloat[]
{
\includegraphics[height=4.0 cm, width=2.4 cm, clip=true, angle=-90]
{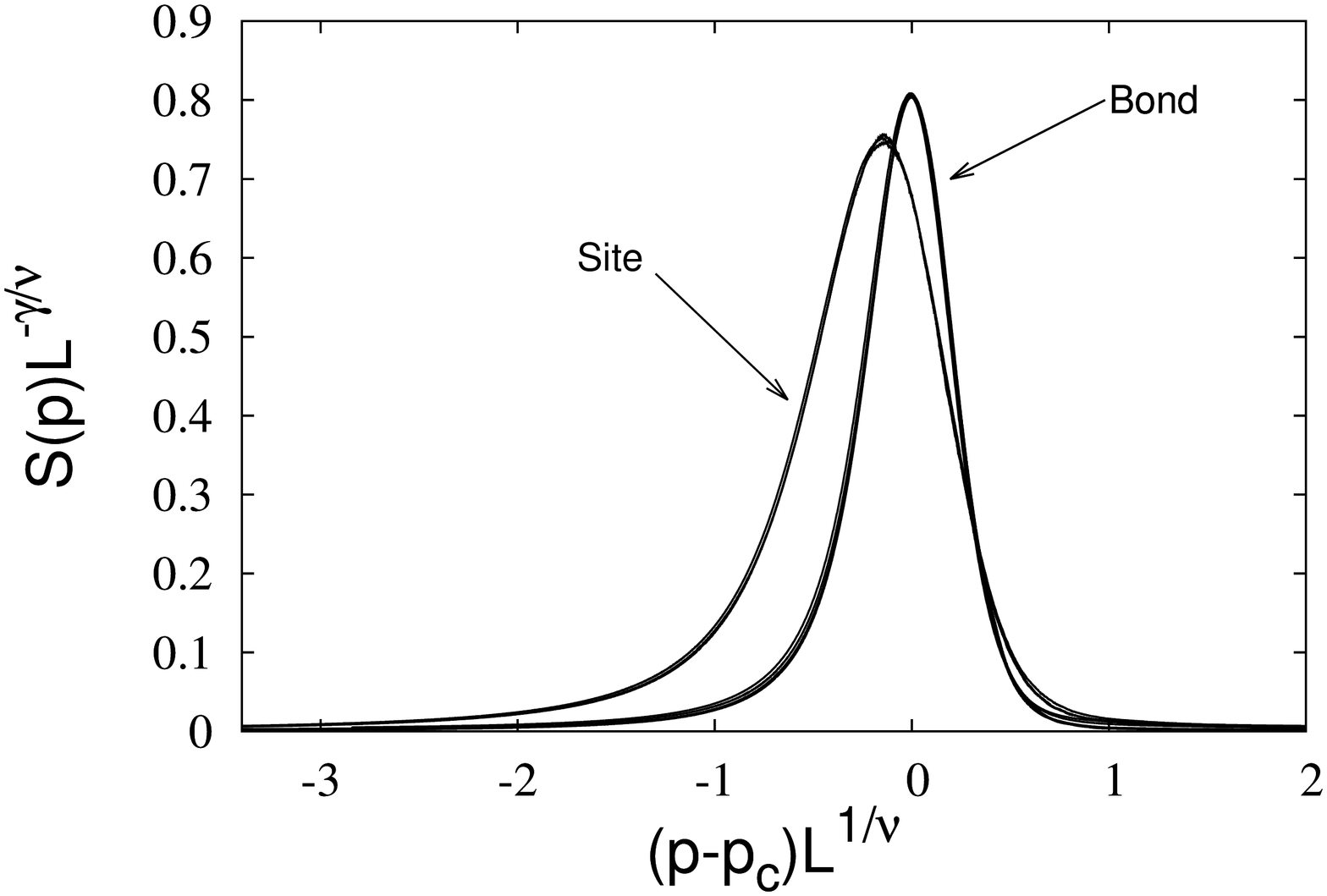}
\label{fig:4d}
}

\caption{The mean cluster size $S(p,L)$ for (a) bond and (b) site percolation as a function of $p$ for different size of the WPSL. 
In the case of bond the cluster size is measured by the number of sites each cluster contain and
in the case of sites it is the area of the contiguous blocks that belong to the same cluster. In (c) we plot $\log S$ vs $\log L$ using the size of $S$ for fixed value of $(p-p_c)L^{1/\nu}$ and find
almost parallel lines with slope $\gamma/\nu=1.73153\pm -0.001979$ and $1.72806 \pm 0.001993$ for bond and site respectively.
In order to obtain a better estimate for the $\gamma$ value we also plot the same data of (a) and (b) in the self-similar coordinates namely
$PS^{-\gamma/\nu}$ and $(p-p_c)L^{1/\nu}$. By tuning the $\gamma/\nu=1.728$ value we find a set of excellent data-collapse for both (a)
and (b) that gives $\gamma=2.825$.} 
\label{fig:4abcd}
\end{figure}

The cluster size distribution function $n_s(p)$ plays a central
role in the description of percolation theory. It is defined as the number of clusters of size $s$ per site
in the lattice. Unfortunately, only in the case of one dimensional system, we know an exact form for the cluster number 
$n_s(p)$ and manage to it handle approximately for infinite system which is actually the Bethe lattice. 
For $1<d<\infty$ we do not yet know an exact expression for $n_s$. 
This is because in such cases there exists a large number of different ways in which clusters of same
size can arrange themselves, which are called lattice animals. 
Even for relatively small cluster size in 
the square lattice we run into difficulties in enumerating them. Nevertheless, theoretically
we can still write down the general expression
\begin{equation}
\label{eq:nsp1}
n_s(p)=\sum_{s,t}g_{s,t}p^s(1-p)^t,
\end{equation}
where $g_{s,t}$ is the number of possible lattice configurations of size $s$ and perimeter of size $t$. 
Note that the quantity $sn_s(p)$ is the probability that an arbitrary site belongs to a cluster of size $s$. On the other hand,  
the quantity $\sum_{s=1} sn_s$ is the probability that an arbitrary site belongs to a cluster 
of any size which is in fact equal to $p$. Therefore, the ratio of the two
\begin{equation}
\label{eq:nsp2}
f_s={{sn_s(p)}\over{\sum_{s=1} sn_s}},
\end{equation}
is the probability that an occupied site chosen at random belong to a cluster of size exactly equal to $s$.
The mean cluster size $S(p)$ therefore is given by
\begin{equation}
\label{eq:nsp3}
S(p)=\sum_s sf_s={{\sum_s s^2n_s}\over{\sum_s sn_s}},
\end{equation}
where the sum is over the finite clusters only i.e., the spanning cluster is excluded from the enumeration
of $S$. The definition of mean cluster size $S$, however, 
does not have information about the geometric structure of the clusters like 
their compactness and spatial extent.
It is important to mention that the mean area of the blocks in the WPSL decreases as $(1+3t)^{-1}$ and hence increasing the size of the lattice 
we need to blow up the lattice by a factor of $3t$. It compensates the decreasing block size with increasing block number $N$. That is, the mean cluster size 
\begin {equation}
\label{eq:nsp4}
S={{1}\over{p}}\sum_s s^2n_s \times 3t,
\end{equation}
in the case of WPSL. In the case of bond percolation, however, we do not
need to multiply by the factor $3t$ as the cluster size here 
is measured by the number of nodes or vertices it contains not by the area.

In Figs. (\ref{fig:4a}) and (\ref{fig:4b}) we show the plots of the mean cluster size $S(p)$, for both bond and site percolation, as a function of $p$ for  different lattice sizes $L$. 
We observe that in either cases, there are two main effects as we
increase the lattice size. First, we see that the mean cluster size
increases  as we increase the occupation probability till $p$ approaches to $p_c$ and the peak height grows profoundly with $L$ in the vicinity of $p_c$. Second, there is a slight shift in the peak towards 
$p_c$ value as we increase $L$. The extent of shift is again given by 
Eq. (\ref{eq:sp1}). To bring the peak height to meet at the same point we first plot $S$ as
a function of dimensionless quantity $(p_c-p)L^{1/\nu}$. 
We then measure the peak height for a fixed value of $(p_c-p)L^{1/\nu}$ but for different $L$. 
Plotting these peak heights as
a function of $L$ in the $\log$-$\log$ scale give straight lines for site and bond percolation both 
(see the inset of Fig. (\ref{fig:4c}). It implies that 
\begin{equation}
\label{eq:nsp5}
S\sim L^{\theta},
\end{equation}
where like before we again choose $\theta=\gamma/\nu$ for future convenience and find that $\gamma/\nu=1.73153 \pm 0.001979$ for bond and $1.72806 \pm 0.001993$ for site. The two values are so close that they
can be well approximated to be the same.
Plotting now the same
data of Figs (\ref{fig:4a}) and (\ref{fig:4b}) by measuring the mean cluster size
$S$ in unit of $L^\theta$ and $(p_c-p)$ in unit of $L^{-1/\nu}$ respectively we find that all the distinct plots of $S$ vs $p$ collapse superbly into one universal curve (see Fig. (\ref{fig:4d})) in both
cases with the same value for the corresponding exponents $\gamma/\nu=1.728$. 
It again implies that the mean cluster size too, for both bond and site, exhibits finite-size scaling
\begin{equation}
\label{eq:nsp6}
S \sim L^{\gamma/\nu}\phi \Big ((p_c-p)L^{1/\nu}\Big ),
\end{equation} 
sharing the same critical exponents. 
Eliminating $L$ from Eq. (\ref{eq:sp1}) in favor of $(p_c-p)$ using $(p_c-p)\sim L^{-1/\nu}$ we find that the mean cluster diverges 
\begin{equation}
\label{eq:nsp7}
S\sim (p_c-p)^{-\gamma},
\end{equation}
where $\gamma=2.825$ for both site and bond percolation. This value is significantly different from the 
known value $\gamma= 2.389$ for all the regular planar lattices.

\begin{figure}[htb]
\includegraphics[width=5.0cm,height=8.5cm,clip=true,angle=-90]{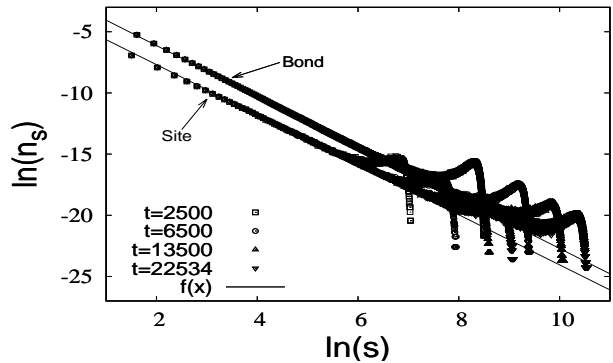}
\caption{We plot the cluster size distribution function $\log(n_s(p_c))$ vs $\log s$ for different 
size of the WPSL. Once again we find almost parallel lines since slopes are $2.07252$ and $2.0728$ for bond and site percolation respectively which implies that the $\tau$ value is independent of the type of percolation.
}
\label{fig:nsp}
\end{figure}

The mean cluster size $S$ according to Eq. (\ref{eq:nsp7}) thus diverges as we approach to the 
threshold value $p_c$ as expected. On the other hand, $S$ can diverge, according to Eq. (\ref{eq:nsp3}),
if $\sum_s s^2n_s$ diverges since denominator $\sum_s sn_s$ in the same limit reaches to a constant $p_c$. 
Generally, we know that 
\begin{equation}
\label{eq:rw1}
\sum_{s=1}^\infty s^\alpha= \left \{ \begin{array}{r@{\qquad\qquad}l}                  
{\rm convergent} \hfill  &  {\rm ;} \hspace{0.5cm}{\rm for} \hspace{.4cm} \alpha<-1 \\
{\rm divergent} \hfill & {\rm ;}\hspace{0.5cm} {\rm if} \hspace{.4cm} \alpha\geq -1, \\
\end{array} \right.. 
\end{equation}
and hence we can use it to find out under what condition the numerator of
Eq. (\ref{eq:nsp3}) diverges. It is convenient to assume 
\begin{equation}
\label{eq:nsp8}
n_s(p)\sim s^{-\tau}\phi((p-p_c)^{1/\sigma}s),
\end{equation}
which means $n_s(p_c)\sim s^{-\tau}$ and hence
\begin{equation}
\label{eq:nsp9}
\sum_{s=1}^\infty s^2n_s(p_c)\sim \sum_{s=1}^\infty s^{2-\tau}.
\end{equation}
It implies that $S$ would diverge as $p\rightarrow p_c$ if $(2-\tau)\geq -1$ or $\tau\leq 3$. On the
other hand, we also demand $p_c\sim \sum_{s=1}^\infty sn_s(p_c)$ implies $\sum_{s=1}^\infty sn_s(p_c) 
\sim \sum_{s=1}^\infty s^{1-\tau}$. It implies that $(1-\tau)<-1$ or $\tau>2$. Putting the two constraints together we find that 
$\tau$ must satisfy the bound $2<\tau\leq 3$. We can thus write that 
\begin{equation}
\label{eq:tau}
n_s(p_c)\sim s^{-\tau}
\end{equation}
where $\tau$ is called the Fisher exponent.
We can obtain the exponent $\tau$ by plotting the cluster area distribution function $n_s(p)$ at $p_c$. In Fig. (\ref{fig:nsp})
we plot $n_s(p_c)$ vs $s$, for both site and bond, in the log-log scale and find two parallel lines except near the tail 
where there is a  hump due to finite size effect. However, we also observe that 
as the lattice size $L$ increases the extent up to which we get a straight line increases too. It implies
that if the size $L$ were infinitely large, we would have a perfect straight line obeying
Eq. (\ref{eq:tau}). The slopes of the lines are $\tau=2.07252$ for bond and $\tau=2.0728$ for site. It implies that the exponent $\tau$ is
almost the same $\tau\sim 2.072$ for both site and bond percolation on WPSL and its value is
different than the known value for all known planar lattices $\tau=2.0549$.

\begin{figure}
\centering

\subfloat[]
{
\includegraphics[height=5.5 cm, width=4.0 cm, clip=true]
{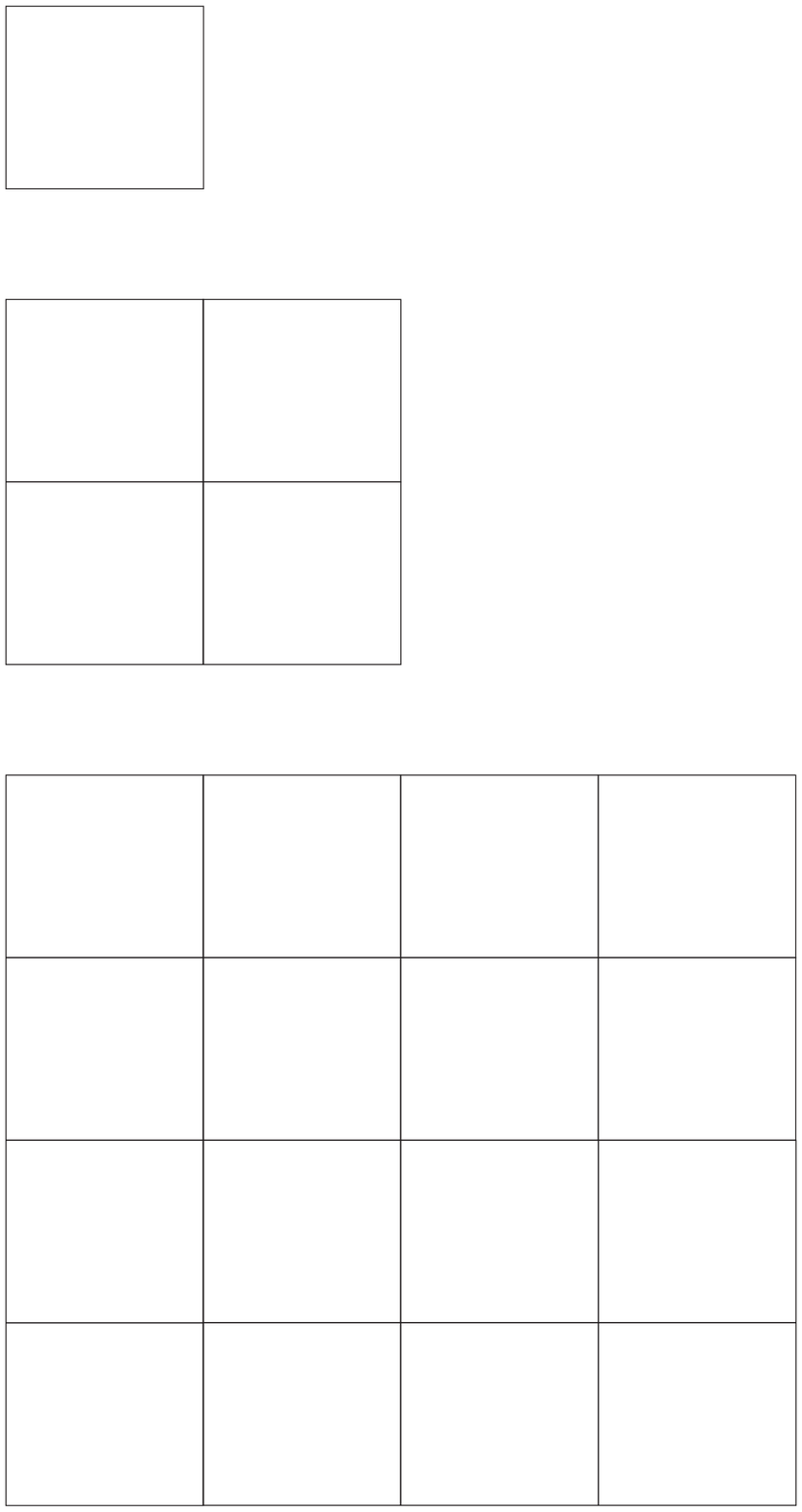}
\label{fig:5a}
}
\subfloat[]
{
\includegraphics[height=5.5 cm, width=4.0 cm, clip=true]
{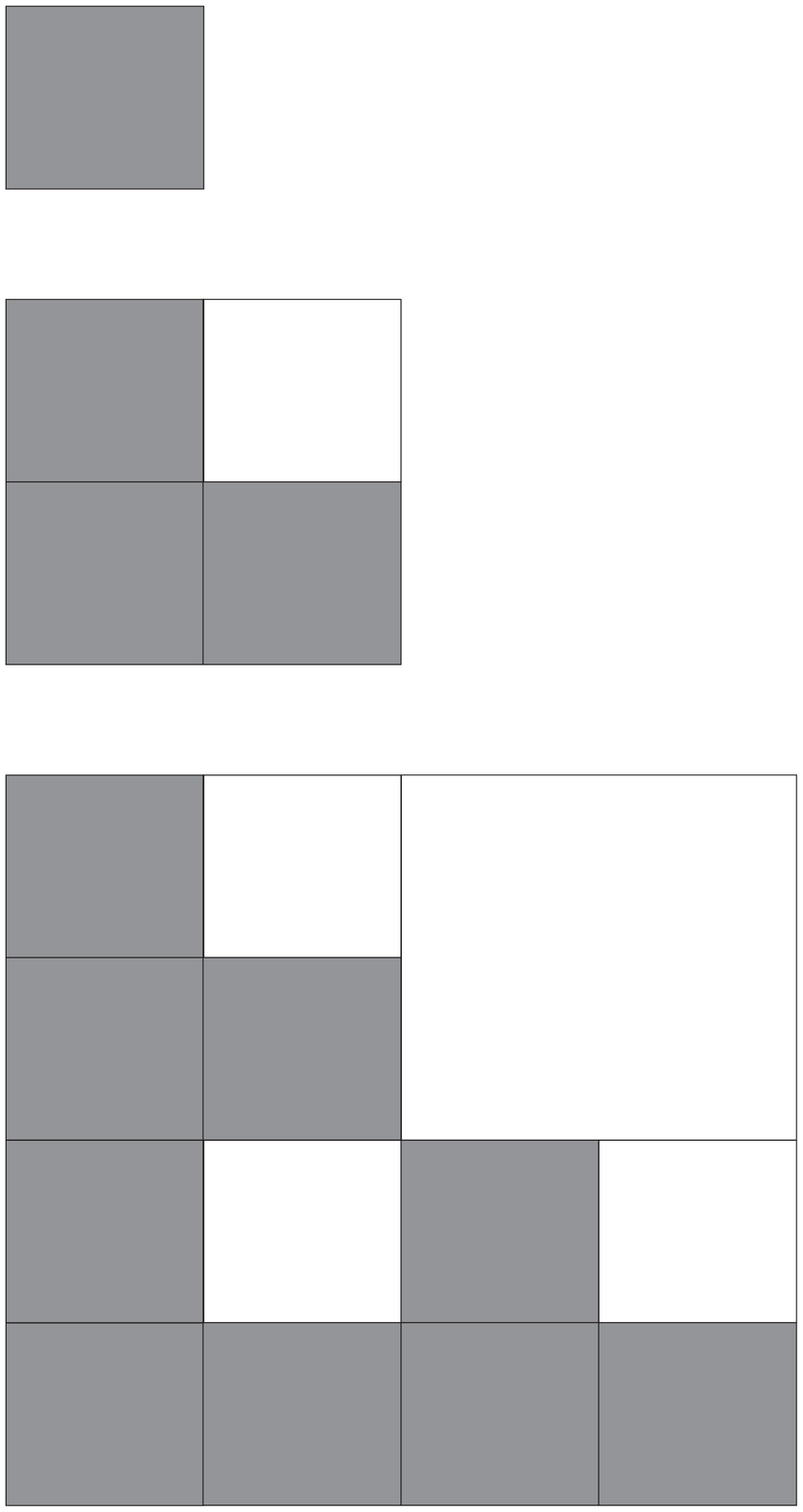}
\label{fig:5b}
}

\caption{Illustration of mass-length relation $M\sim L^D$. (a) Compact or uniform distribution of mass 
gives $D=2$ that coincide with the dimension of the space where it is embedded. It is thus Euclidean is 
nature. (b) The distribution of the same mass gives a highly ramified structure with $D=\ln 3/\ln 2$ which is less than the dimension of the space where it is embedded. It is thus a fractal.} 
\label{fig:ab}
\end{figure}

\begin{figure}[htb]
\centering
\includegraphics[width=5.0cm,height=8.5cm,clip=true,angle=-90]{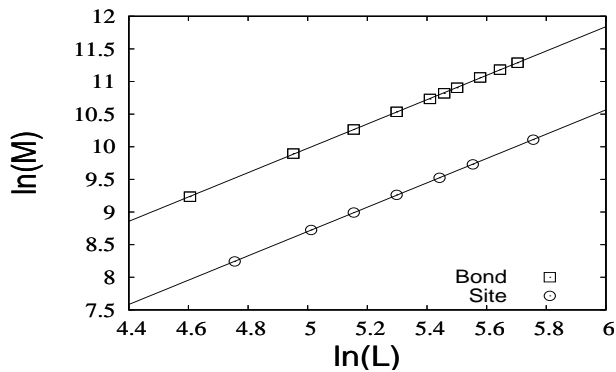}
\caption{The mass of the spanning cluster $M$, the total area in the case of site and the number of sites in the case of bond, is shown as a function of system size $L$ for both site and bond percolation.
The two lines with slope $d_f=1.86378 \pm 0.02249$ for bond and $1.86439 \pm 0.001498$ for site reveals that the fractal dimension of the spanning cluster is independent of the type of percolation.}
\label{fig:fractal}
\end{figure}

Let $M(L)$ denote the mass or size of the percolating cluster of lattice of linear size $L$.
If the percolating cluster grows as a compact object, then its mass $M(L)$ would grow with $L$ 
as $M(L)\sim L^2$ since the dimension of the embedding space of the WPSL is $d=2$. 
However, at $p_c$ if we would like to walk through the spanning cluster then the amount of time
it would take must diverge
as $L\rightarrow \infty$. This is so because of the fact that percolating cluster at $p_c$ 
is highly ramified. 
In fact, if we had $p=1$ that would surely be $M(L)\sim L^2$. At $p_c$ we also get the same mass-length
relation but the exponent is less than $2$. To understand the significance of it, let us
stack objects of unit sized squares as shown in Fig. (\ref{fig:5a}). In step one, we make four copies of unit square. Then we stack two of them side by side and the other two on top of those two also side by side.
In step two, we make four copies of the resulting object after step one. We stack two of them side by side like step one and the other two on top of them
again side by side. In general in the step $i$ we make 
four copies of the resulting object after step $(i-1)$. We then stack two them side by side and the 
other two on top of this two again side by side as shown in Fig. (\ref{fig:5a}). It is easy to check that it obeys
the mass of the object  grow according to the following mass-length relation
\begin{equation}
\label{eq:masslength}
M\sim L^D,
\end{equation}
with $D=2$. Now, let us slightly change the situation. We do everything like before with the only difference
is that at each step we throw the top right copy leaving its space empty as shown in Fig. (\ref{fig:5b}). The amount of mass of the resulting system in the $i$th step is $M=3^i$ and the linear size of the system is $L=2^i$. 
Using this two relations we can eliminate $i$ in favor 
of $L$ and we find
the same mass-length relation as in Eq. (\ref{eq:masslength}) except that we get exponent $D=\ln 3/\ln 2$ \cite{ref.multifractal_1}. We could even remove any of the four 
copies at random and still we would get the same result. 
The exponent of the mass-length relation $D=d_f$ which is now less than the dimension of the embedding space $d=2$ and hence
it is a fractal. The spanning cluster too is highly ramified like Fig. (\ref{fig:5b}) as it has holes of many different sizes. Now, a litmus test whether the spanning cluster is a fractal or not would be to check if 
it obeys the same mass-length relation with an exponent $d_f<2$ since the embedding space of the spanning cluster is a plane. We plot the size of the spanning cluster $M$ as a function of lattice size 
$L$ in the log-log scale as shown in Fig. (\ref{fig:fractal}). Indeed, we find that $d_f=1.86439\pm 0.001498$ for site and $1.86378 \pm -0.02249$ which are almost the same but significantly
different from the one for regular planar lattices $d_f= 1.895$. It may appear that the difference between the $d_f$ for WPSL and that for regular planar lattices is not much
but it important to remember that even a small difference in fractal dimension has an huge impact in their degree of ramification.

We already know that the mean
cluster size diverges i.e., $S\rightarrow \infty$ as $p\rightarrow p_c$. 
According to Eq. (\ref{eq:nsp3}), $S$ can only diverge
if its numerator diverges. Generally, we know that $\sum_{s=1}^\infty s^\alpha$ converges if $\alpha<-1$ and diverges if 
$\alpha\geq -1$. Applying it into both numerator and denominator of Eq. (\ref{eq:nsp3}) at $p_c$ gives a bound that $2<\tau<3$. Using Eq. (\ref{eq:nsp8}) in Eq. (\ref{eq:nsp3}) and taking continuum limit gives
\begin{equation}
S\sim s_\xi^{3-\tau}.
\end{equation}
We know that $s_\xi$ diverges like $(p_c-p)^{-1/\sigma}$ where $\sigma=1/(\nu d_f)$ and hence comparing it with Eq. (\ref{eq:nsp7})
we get
\begin{equation}
\label{eq:tau_scaling}
\tau=3-\gamma \sigma.
\end{equation}
 Besides, there is another well known scaling relation $\tau=1+d/d_f$ which we can use to find $\tau$ value. Using the $d_f$ value for WPSL
in the scaling relations, $\tau=3-\gamma \sigma$ and $\tau=1+d/d_f$, we find $\tau$ equal to 
$2.0725$ and $2.0728$ respectively which is almost equal to the one we obtained straight from slope of Fig. (\ref{fig:fractal}). 
There are also a couple of other well-known scaling relations, such as  
$\beta=\nu(d-d_f)$, $\gamma=\nu(2d_f-d)$, which we used for a consistency check of our results. To this 
end, we find that our estimates satisfy these relations up to quite a good extent. 

\begin{table}[h!]
\centering
    \begin{tabular}{| l | l | l |}
    \hline
    Exponents & regular 2d lattice & WPSL bond/site \\ \hline
    $\nu$ & 1.75 & 1.635  \\ \hline
    $\beta$ & 0.13889 & 0.222  \\ \hline
    $\gamma$ & 2.3889 & 2.825  \\ \hline
   $ \tau$ & 2.0549 & 2.0728 \\ \hline
$d_f$ & 1.895(8) & 1.864 \\

    \hline
    \end{tabular}
\caption{The critical and other characteristic exponents for site and bond percolation in the WPSL and
in the regular planar lattice are given alongside.}
\label{table:1}
\end{table}

\section{Summary and discussion}

In this article, we have studied both bond and site percolation on WPSL using extensive Monte Carlo simulations. 
We thought it is important to know some key features of the WPSL so that one can understand
why it is so special and unique. We therefore have first
briefly discussed its construction process and then its various properties which are as follows. 
(i) The dynamics of
its growth is governed by infinitely many conservation laws. (ii) Its area size 
distribution function obeys dynamic scaling.
(iii) Each of the infinitely many conservation laws, except conservation of total area, gives rise
to multifractal spectrum and hence WPSL is a multi-multifractal. Fourth, its coordination number distribution
function follows a power-law. (iv) It has a mixture of properties of both lattice and graph.
On one hand, like lattice, it is embedded in a space of dimension $D=2$; On the other its coordination number distribution follow power-law like network. These unique properties
have resulted in unique results too. We also briefly discussed about the finite-size scaling
theory and have shown that its origin is deeply rooted to the Buckingham $\Pi$-theorem. The finite-size
scaling is one of the most crucial aspects in percolation as it helps extrapolating critical exponents
for infinite system using data for a set of finite size systems. This is done by using the idea of data 
collapse. Note that an excellent data collapse is one of the clear testaments that the numerical
values we obtained for various exponents are quite satisfactory. Besides, we show that these  
satisfy a set of scaling relations which also provide a consistecy check.

In this work we first obtained
percolation threshold $p_c=0.3457$ and $p_c=0.5265$ for bond and site percolation on WPSL. Naturally, the $p_c$ for bond is less 
than that of its site counterpart as expected. We also obtained numerically the various observable quantities
such as the spanning probability $W(p)$, the percolation strength $P(p)$, the mean cluster size $S(p)$
etc. using NZ algorithm. The initial data obtained from the NZ algorithm correspond to microcanonical ensemble. To get the corresponding data that correspond to canonical ensemble we used the convolution equation given
by Eq. (\ref{eq:convolution}) for each observable quantities. 
With the help of a comprehensive finite-size scaling theory 
we also obtained numerically the critical exponents $\nu, \beta$ and $\gamma$ for both
bond and site percolation on WPSL and confirm they are equal (see table \ref{table:1} for detailed comparison). To check further if they are equal or not we used the idea
of data collapse and found an excellent data collapse for the same critical exponents albeit different
$p_c$. Note that good estimate of $p_c$ and of the critical exponents a must for obtaining satisfactory
data collapse. These 
values also satisfy the scaling relations. All these provide a clear testament that
the critical exponents for bond and site percolation in WPSL are the same. It happens in spite of the
significant difference in the definition of clusters. Interestingly,
these values are significantly different from the ones for all known planar lattices. 
We can thus conclude that
the universality class of WPSL (bond and site) is distinct from the ones for all the known planar lattices.
It happens in spite of the significant differences in the definition of site and bond 
in the WPSL.

Hsu and Huang also studied percolation in a class of random planar
lattices and their duals yet they found the same critical exponents as the ones for regular lattices.  Corso {\it et. al.}
studied percolation on multifractal planar lattices and they too found the same critical exponents as the ones on regular lattices. So, it is neither the randomness nature of the lattice
nor the multifractal nature of the lattice can be held responsible for making WPSL unique. The planar random lattice that Hsu and Huang studied is quite different than WPSL. The coordination 
number distribution of their lattice do not obey power-law. This is perhaps one of the most significant differences. Or, it may be the case that when a lattice is multifractal and at the
same time it is random then depending on further detailed nature may be responsible for giving a new set of exponents. However, it is too soon to draw any conclusion. 
We hope to device more variants of WPSL in our future endeavour and see what happens. Nevertheless, we still hope that our findings will
have a significant impact in the percolation theory.

\end{document}